\documentclass[twocolumn]{aastex63}
\usepackage{graphicx}
\usepackage{amsmath}
\usepackage{amssymb}
\usepackage{wasysym}
\usepackage{color}

\begin{document}

\title{The Flattening Metallicity Gradient in the Milky Way’s Thin Disk}
\shorttitle{Flattening MW [Fe/H] Gradient Over Time}
\shortauthors{Vickers et al.}

\author[0000-0001-6049-8929]{John J. Vickers}
\affiliation{Department of Astronomy, School of Physics and Astronomy, Shanghai Jiao Tong University, 800 Dongchuan Road, Shanghai 200240, People's Republic of China}
\affiliation{Key Laboratory for Particle Astrophysics and Cosmology (MOE) / Shanghai Key Laboratory for Particle Physics and Cosmology, Shanghai 200240, China}
\email{johnjvickers@sjtu.edu.cn, johnjvickers@gmail.com}

\author[0000-0001-5604-1643]{Juntai Shen}
\affiliation{Department of Astronomy, School of Physics and Astronomy, Shanghai Jiao Tong University, 800 Dongchuan Road, Shanghai 200240, People's Republic of China}
\affiliation{Key Laboratory for Particle Astrophysics and Cosmology (MOE) / Shanghai Key Laboratory for Particle Physics and Cosmology, Shanghai 200240, China}
\affiliation{Key Laboratory for Research in Galaxies and Cosmology, Shanghai Astronomical Observatory, Chinese Academy of Sciences, 80 Nandan Road, Shanghai 200030, People's Republic of China}
\email{jtshen@sjtu.edu.cn}

\author[0000-0001-5017-7021]{Zhao-Yu Li}
\affiliation{Department of Astronomy, School of Physics and Astronomy, Shanghai Jiao Tong University, 800 Dongchuan Road, Shanghai 200240, People's Republic of China}
\affiliation{Key Laboratory for Particle Astrophysics and Cosmology (MOE) / Shanghai Key Laboratory for Particle Physics and Cosmology, Shanghai 200240, China}
\email{lizy.astro@sjtu.edu.cn}

\def\mean#1{\left< #1 \right>}

\def\zmax{$Z_{max}$}
\def\rguide{$R_{\mathrm{g}}$}
\def\rpres{$R_{present}$}

\renewcommand{\thempfootnote}{\arabic{mpfootnote}}

\begin{abstract}

  We calculate the ages, orbits and phase-space coordinates for a sample of $\sim$4 million LAMOST and Gaia stars. The ages are crossmatched and compared with values from two other popular age catalogs which derive the ages using different methods. Of these $\sim$4 million stars, we select a sample of 1.3 million stars and investigate their radial metallicity gradients (as determined by orbital radii) as a function of their ages. This analysis is performed on various subsets of the data split by chemistry and orbital parameters. We find that commonly used selections for ``thin disk'' stars (such as low-$\alpha$ chemistry or vertically thin orbits) yield radial metallicity gradients which generally grow shallower for the oldest stars. We interpret this as a hallmark feature of radial migration (churning). Constraining our sample to very small orbital \zmax \ (the maximal height of a star's integrated orbit) makes this trend most pronounced. A chemistry-based ``thin disk'' selection of $\alpha$-poor stars displays the same trend, but to a lesser extent. Intruigingly, we find that ``thick disk'' selections in chemistry and \zmax \ reveal a slightly positive radial metallicity gradients which seem similar in magnitude at all ages. This may imply that the thick disk population is well mixed in age, but not in radius. This finding could help constrain conditions during the early epochs of Milky Way formation, and shed light on processes such as the accretion and reaccretion of gasses of different metallicities.

\end{abstract}

\section{Introduction}\label{sec:introduction}

Stellar radial migration is a phenomenon by which a star may find itself at a Galactocentric radius far from the radius where it was born. Typically, there are two major mechanisms related to radial displacement:
\begin{itemize}
\item \emph{Churning} is the process by which a star exchanges angular momentum with non-axisymmetric features in the potential and is subsequently moved from one mean orbital radius to another. Transient spiral arms are the most commonly discussed feature associated with this process (\citealt{sel2002}; see further work by \citealt{ros2008}, \citealt{ros2012}).
\item \emph{Blurring} is the process by which a star's present day position is moved away from its orbital position through epicyclic motion, causing metallicity gradients to be blurred out and flattened \citep{sel2014}. It is commonly associated with heating processes such as collisional interactions with gas clouds, satellites, or other stars. (\citealt{spi1951}, \citealt{lac1984}, \citealt{aum2016}). Technically blurring is not migration, since the mean orbital radius of the star remains unchanged by this process.
\end{itemize}
  
Churning is thought to be most influential on stars which orbit the galaxy on cold, nearly circular tracks. This is because these stars, with fewer orbital frequencies, remain resonant longer when captured. As a side-effect, it is thought that younger stars may be more prone to large migrations while older stars are more resistant; although this is more a function of orbital vertical excursion magnitude, or orbital circularity, than age (the so-called ``provenance bias;'' \citealt{ver2014}, \citealt{ver2016}; see also \citealt{mik2020}). \citet{sol2012} note that thick disk stars may have their angular momenta changed as much as thin disk stars when exposed to a realistic potential with spiral arms; they also note that the stars which migrate the most generally have the most circular orbits and smallest orbital scale heights. Churning works to flatten the metallicity gradient in guiding radius (\rguide) as stars' orbital radii are changed away from their birth locations.

Blurring is thought to grow fairly constantly over time as the number of interactions should increase steadily with time for an object orbiting in the vicinity of the disk. This leads to the well studied Age-Velocity-Dispersion relation (\citealt{str1946}, \citealt{hol2009}, \citealt{vic2018}, \citealt{mac2019}) and contributes to the flattening of metallicity gradients in \rpres \ over time as stars from different radii mix in observational space and ``blur'' the original metallicity gradients of the interstellar medium (\citealt{edv1993}, \citealt{sch2009}). 

Churning and blurring operate simultaneously in the Galaxy, with the effect of churning perhaps being an order of magnitude larger than that of blurring \citep{fra2020}. This leads to churning flattening the metallicity gradient much more significantly than blurring over time.

This flattening of the metallicity gradient over time provides an interesting and inverted way to directly measure and quantify the effects of churning. If we assume that the Galaxy formed in an inside-out fashion (\citealt{mat1989}, \citealt{bir2013}), then older stars should have a steep metallicity gradient and younger stars should have a shallower one (as the gradient in the interstellar medium is commonly believed to flatten out over time; see, for example: \citealt{min2018} and cosmological simulations presented in \citealt{pil2012}, \citealt{gib2013}). If we do not observe this pattern in the present day, migration is a likely culprit.

The metallicity gradient as a function of age has been investigated in a variety of datasets using a variety of methods (for example: \citealt{nor2004}, \citealt{cas2011}, \citealt{yu2012}, \citealt{xia2015}, \citealt{wan2019}).

In this work, we aim to improve on past studies by leveraging the massive LAMOST-Gaia overlapping data set, which enables us to calculate higher precision isochrone ages than have generally been possible before Gaia (thanks to Gaia parallaxes unlocking absolute magnitude space) for $\sim$1.3 million stars. These improvements in the quality and, most notably, size of our data allow us to investigate the metallicity gradients for a variety of slices in chemical and orbital parameter spaces extending from thin-disk like selections to thick disk ones, offering a comprehensive view of the effect. We will also investigate how the gradients and trends change when using different assumptions of Galactic parameters such as the potential and Solar characteristics.

In Section \ref{sec:data}, we discuss our data set, the calculation and verification of age determinations, phase-space coordinate assignments, and orbital integrations. In Section \ref{sec:results} we outline our primary findings related to the metallicity gradient in thin and thick disk like populations, and we detail how these findings may change with different experimental choices in Section \ref{sec:discussion}. We conclude in Section \ref{sec:conclusions}.

\section{Data}\label{sec:data}
\subsection{Observational Data}\label{ssec:obsdata}

For this work we will be relying primarily on age and kinematic information, and so will be using spectroscopic and astrometric data.

The spectroscopic data is provided by the Large Area Multi Object Spectroscopic Telescope (LAMOST; \citealt{cui2012}, \citealt{luo2015}) survey. LAMOST is a Schmidt Telescope with a 4-meter primary reflector and a five degree field of view. The telescope is guided by four CCDs embedded in the focal plane and the primary science readings come from 4000 spectroscopic fibers which position themselves with the aid of two small motors each. The spectra we use are collected at a resolution of R$\sim$1800 in the visual wavelengths and extend from about 10 to 18 in $g$ magnitude (for details about the observing strategy, see \citealt{den2012}, \citealt{yua2015} and related references). We utilize DR5, which contains approximately 9 million individual spectra, many of which are useful repeat observations. The observations cover the sky between about -10$^{\circ}$ and 60$^{\circ}$ in declination, with specific fields being prioritized like the Kepler field (\citealt{bor2010}, \citealt{fu2020}) and the Galactic anticenter \citep{liu2014}.
  
The LAMOST stellar parameter pipeline \citep{wu2014} supplies atmospheric parameters, but unfortunately does not calculate $\alpha$ abundances as part of the primary data product. Therefore, we use the atmospheric parameters ([Fe/H], [$\alpha$/Fe], T$_{eff.}$, log(g)) provided by the DD-Payne \citep{xia2019} supplementary catalogue, which includes $\alpha$ abundances. The DD-Payne pipeline is a data-driven neural network inspired by \emph{The Payne} \citep{tin2019} and \emph{The Cannon} \citep{nes2015} which is trained on the separate, overlapping subsamples of LAMOST+APOGEE \citep{maj2017}, and LAMOST+GALAH \citep{des2015}.

The errors on our data, for signal-to-noise ratio greater than 50, are:  $\delta$T$_{eff}\sim$22.6 K, $\delta$log(g)$\sim$0.05 dex, $\delta$[Fe/H]$\sim$0.03 dex, $\delta$[$\alpha$/Fe]$\sim$0.02 dex.

Astrometric data comes from the Gaia survey data release eDR3 \citep{gai2020}. This ongoing survey has, to date, collected astrometric and photometric data on 1.8 billion sources with magnitude brighter than 21 in $g$. Parallax, proper motion, and color information is available for 1.5 billion of those sources.

\begin{figure}%[!htb]
	\includegraphics[width=\linewidth]{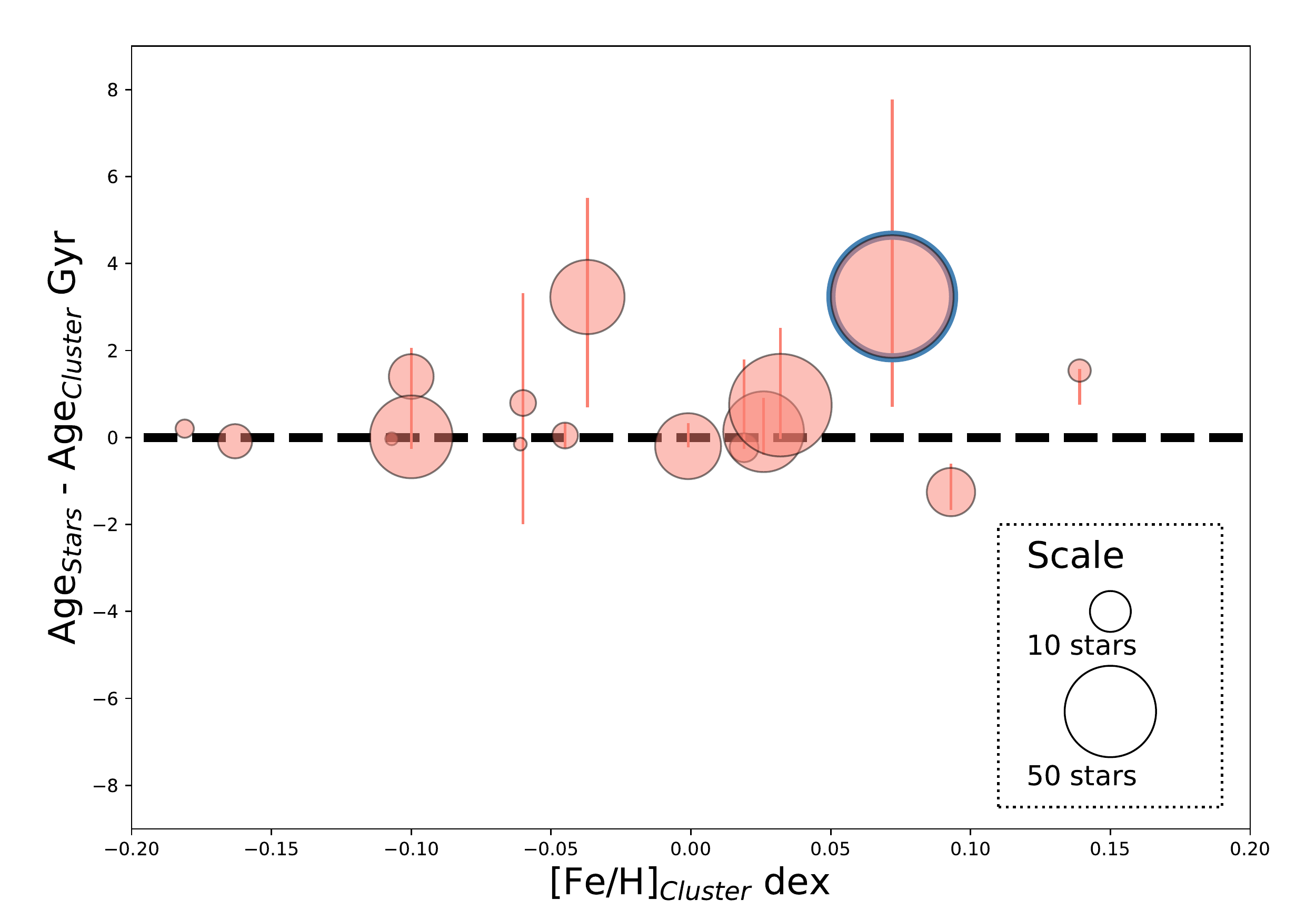}
	\caption{
          The difference in ages between the literature values for a selection of open clusters and the derived ages for members of those clusters as a function of cluster metallicity. The sizes of the circles indicate the number of cluster members for which we have age determinations, and the overplotted lines indicate the spreads of those ages. There appears to be no metallicity bias, that is, the offset between the literature ages and the derived ages does not seem to vary strongly as a function of the [Fe/H] abundance. The dark blue outlined cluster is M67, its constituent members are shown, along with representative isochrones, in Figure \ref{fig:m67}. NGC 6791 is omitted from this figure, it lies at [Fe/H]=0.39 and +1.9 Gyr on this plot.
        }
	\label{fig:clusters}
\end{figure}

\subsubsection{Ages}\label{sec:ages}

Ages are calculated as in \citep{vic2018}; using a method similar to \citep{jor2005}. A grid of isochrones is constructed in age, metallicity, and $\alpha$ abundance. The grid is spaced: in increments of 0.25 Gyr from 1 Gyr to 5 Gyr, and 0.5 Gyr thereafter to a maximum of 14.5 Gyr; 0.1 dex in metallicity from -2.4 to +0.5 dex; and 0.2 dex in $\alpha$ abundances from -0.2 to +0.8 (for [Fe/H] $>$ 0 dex, [$\alpha$/Fe] is limited to the range of -0.2 to +0.2).

For this work we have opted to use the Dartmouth library of isochrones (\citealt{dot2007}, \citealt{dot2008}) because it has implemented $\alpha$ abundances and also has Gaia color filters. One downside of the Dartmouth family of isochrones is that it does not extend to very young stellar ages, with the youngest isochrone being 1 Gyr.

Each star is compared to each isochrone, where a $\chi^{2}$ distance is calculated between the star and each line-segment on the isochrone with respect to the observational errors. This $\chi^{2}$ value is converted into a likelihood, which is then weighted by the initial mass function of \citet{kro2001} corresponding to each line segment of the isochrone. The integral of these likelihood values over the full mass range of the isochrone gives the likelihood of the star belonging to that specific isochrone. We then marginalize over all [Fe/H] and [$\alpha$/Fe] values to find the posterior probability of a star being a given age (Equations 1-9 in \citealt{vic2018}, with [$\alpha$/Fe] values accounted for in addition to [Fe/H]).

When comparing magnitudes to isochrones, we use Gaia photometry which has been extinction corrected with the $mwdust$\footnote{https://github.com/jobovy/mwdust} library detailed in \citet{bov2016}. This library determines extinction values at any given coordinate using the 3D reddening maps of \citet{dri2003}, \citet{gre2015}, and \citet{mar2006}.

We use distances and errors from the catalog of \citet{bai2020}. The asymmetric errors in the distance catalog are accounted for by sampling a two-piece Gaussian, that is, a broken Gaussian with different $\sigma$ values, but the same probability mass, on each side of the mean. This preserves the probability mass on both sides of the catalogue median, but suffers discontinuity at the center. An alternative option which we did not use is the ``Fechner'' distribution, which again uses different $\sigma$ values to describe the positive and negative halves of the distribution, but stitches the two Gaussian halves together at the peak. The ``Fechner'' distribution preserves continuity at the mean, but does not preserve the balance of probability mass to the positive and negative directions.

To increase the speed of the calculation, isochrone points are only considered when they are within ten times the error of the stellar parameter in question.

Errors on our age calculations are derived following the procedure outlined in \citet{jor2005}. We first calculate the likelihood of a star being each age in our grid of possible ages (0.25 Gyr, 0.5 Gyr, etc.), we then linearly interpolate the likelihoods between each of these points to create a likelihood surface. Next, we find the \emph{maximal} extent in age such that all likelihoods \emph{outside} of that extent are less than 0.61 times the maximum likelihood (there may be some likelihoods within that extent which are below 0.61 times the maximum likelihood, as in the case where a star is well described by two different age estimates). This cut of 0.61 has been found to well describe the 68\% confidence intervals by \citet{jor2005}.

For those unfamiliar with stellar ages, there are a few things to keep in mind about isochrone stellar ages. The first is that areas of the parameter space where isochrones change dramatically with age (for example the turn-off or sub-giant branch) generally give more precise ages than areas that are similar for all ages (such as the cool main-sequence or the giant branch); this is why some authors carefully select turn-off and sub-giant samples when working with stellar ages. The second is that observed atmospheric parameter errors will naturally cause stars to be scattered outside of the extents of the isochrone grid, which leads to pile-ups of stars in the edge age bins. The third would be that younger isochrones change faster than older isochrones, which means younger stars generally have smaller absolute age errors that older stars (which is why many authors quote age errors as a percentage of age, rather than a Gyr value, or work with logarithmic ages). Some of these quirks will be referenced throughout the paper.

To investigate the accuracy and possible biases in our ages, we compare our ages to those of open clusters. We first select all of our data which have been assigned cluster membership in the catalog of \citet{can2018}, and then compare the ages of these cluster members to the literature ages of the clusters given in the catalog of \citet{dia2021}. There are 17 clusters in our comparison which have more than one member in our data and are present in both these catalogs.

In Figure \ref{fig:clusters} we show the distribution of clusters in metallicity space, and the difference in age between our calculation and the literature values. While there appears to be an offset between our ages and the literature values given for the clusters (ours being about 0.66 Gyr older), there appears to be no significant trend between age offset and metallicity.

It is relatively well known that the Dartmouth family of isochrones ``runs hot'' or, alternatively ``runs old.''
We find that our Dartmouth isochrone ages are about 1.8 Gyr older than the ages derived by \citet{san2018}, who use the PARSEC isochrones (see Section \ref{sec:age_comp}), for example.

\subsubsection{Comparison with M67}\label{sec:m67comp}
Some aspects of the test in Section \ref{sec:ages} prompt closer investigation. For example, in Figure \ref{fig:clusters}, our cluster with the most members and also one of the largest age offsets is NGC 2682 (M67). We see an offset of 3.24 Gyr between the cluster's literature value (3.76 Gyr) and the ages we calculate for the constituent stars (7.0 Gyr). However, the literature value is perhaps not consistent with more recent studies of the cluster, such as that of \citet{bar2016}, which places the cluster's age closer to 4.2 Gyr and thus our offset closer to 2.8 Gyr.

\begin{figure}
	\includegraphics[width=\linewidth]{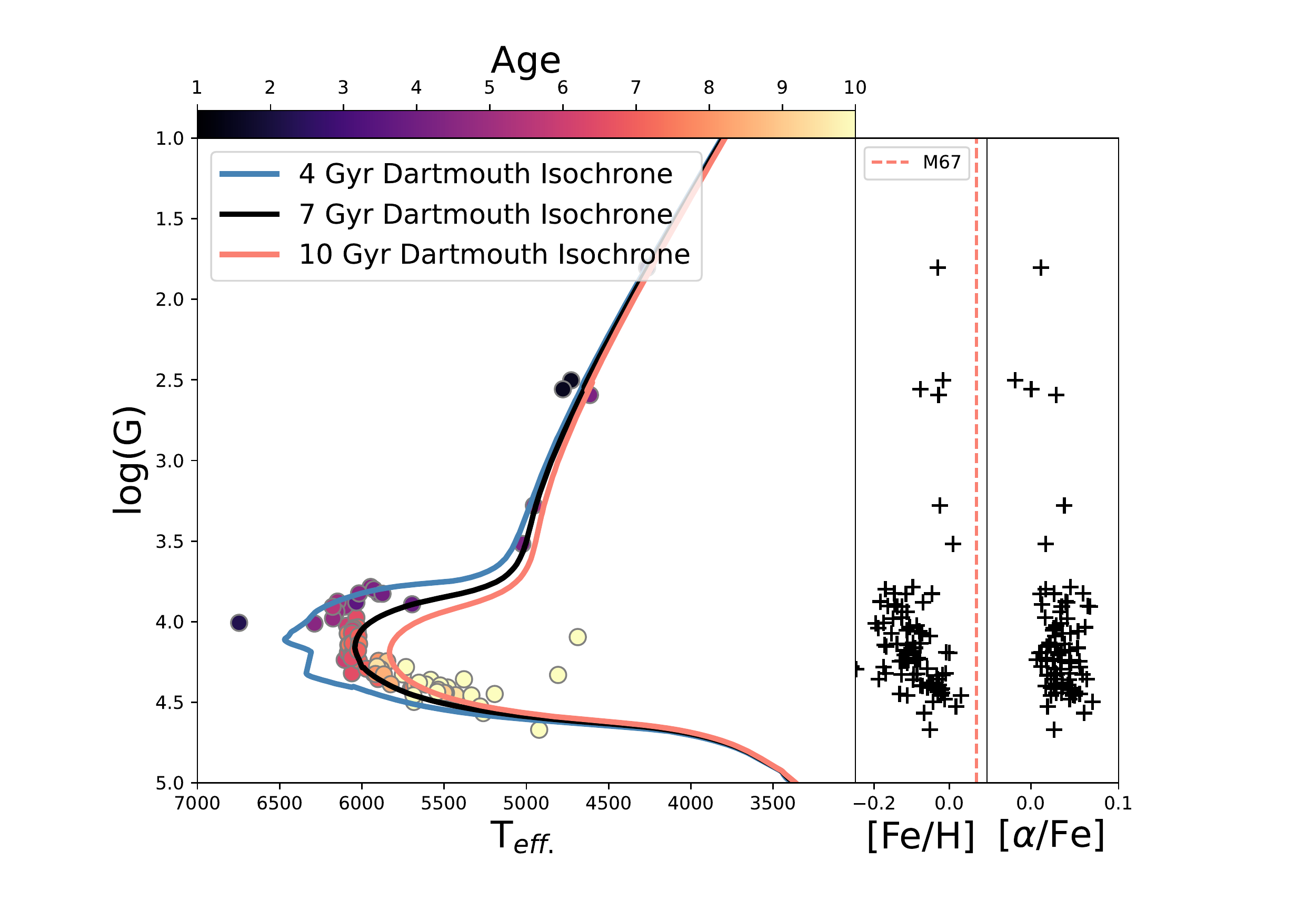}
	\caption{
          The stellar data in M67, our cluster with the most members in our data, and also with one of the largest age offsets from the literature. Three isochrones are overlaid, they all have [Fe/H]=-0.1 and [$\alpha$/Fe]=0.0, which are the values in our isochrone grid nearest the mean values of the member stars. The youngest isochrone (blue), which is approximately equal to the consensus literature age value for this cluster, fits the upper turn-off branch well, but misses the lower turn-off region. The intermediate age isochrone (black) intersects the bulk of the lower turn-off region (where the most stars are in our data). The oldest age isochrone (red) misses most features of the stellar data, except for the main-sequence.
          The panels to the right show the [Fe/H] and [$\alpha$/Fe] values for the cluster members. We see that as we move from the main sequence (at the bottom) to the turnoff and subgiant branch, there is a slight trend toward lower metallicities. This trend then reverses with the giant branch stars being more metal rich than the turnoff and subgiant branch in our data. There appears to be no trend in [$\alpha$/Fe] with surface gravity. The literature value for the [Fe/H] abundance of M67 is shown as a dashed red line in the appropriate panel. $\alpha$ abundances for the cluster are not presented in the catalog we used. Note that three stars are omitted in each panel for visibility of the trends.
        }
	\label{fig:m67}
\end{figure}

We further explore the offset in this particular cluster in Figure \ref{fig:m67}. We plot our 90 member stars, color coded by their age determinations, along with three isochrones which all have [Fe/H]=-0.1 and [$\alpha$/Fe]=0.0, the closest isochrone in our grid to the member stars' values. We see that the literature-age isochrone does not align with the majority of the stars in the lower turn-off and main-sequence region. An intermediate age isochrone overlays the data well, except for the upper turn-off region. An old-age isochrone misses the turn-off entirely.

A pattern is visible in the individual stars, with log(g) appearing correlated with age; although comparison with the overplotted isochrones makes this correlation appear not-unreasonable. Note that this figure does not show absolute magnitude, which is another constraint in the age-fitting procedure.

It is common in stellar age studies to encounter an age-metallicity degeneracy where \emph{younger} ages can be misinterpretted as \emph{lower} metallicities (as both parameters tend to move stars \emph{blueward} or toward \emph{higher} temperatures). This degeneracy could possibly explain the offset we see in age (with our ages being systematically older than the literature values) and [Fe/H] (with our metallicities being systemically lower than the literature values).

Examining our [Fe/H] data more closely, we see that there is a trend in metallicity, in that our data tends to grow more metal poor, and younger, as we travel up the isochrone from the main sequence to the turnoff. This trend is actually the opposite of the expected age-[Fe/H] degeneracy pattern.

These contradictory patterns -- that our stars are too old and too metal poor for the cluster as a whole, and yet as we increase the offset in metallicity within the cluster, the age difference grows smaller -- prevent us from making a simple correction to these data. The topic deserves more attention than we give it here, but unfortunately it seems to be beyond the scope of the current work.

While our data has ages for this cluster which do not match the literature value, we see no obvious reason to distrust our age-calculations in Figure \ref{fig:m67}. Rather, we reiterate to the readers that stellar ages rely on many different complex systems working in tandem and findings relying on them should be considered in that light.

\subsubsection{Comparison with other Age Catalogs}\label{sec:age_comp}

Determination of stellar ages is a challenging problem with many complexities \citep{sod2010}. For the interested readers, we briefly compare our age determinations with those from two other major catalogs: the LAMOST catalog of \citet{xia2017}, and the composite age catalog of \citet{san2018}. Owing to the various quality cuts, procedures, and data sources used in these three different works, we have a sample of about 200,000 stars with ages in all three catalogs (rather than the $\sim$million or more in any given one of them).

\begin{figure*}
	\includegraphics[width=\linewidth]{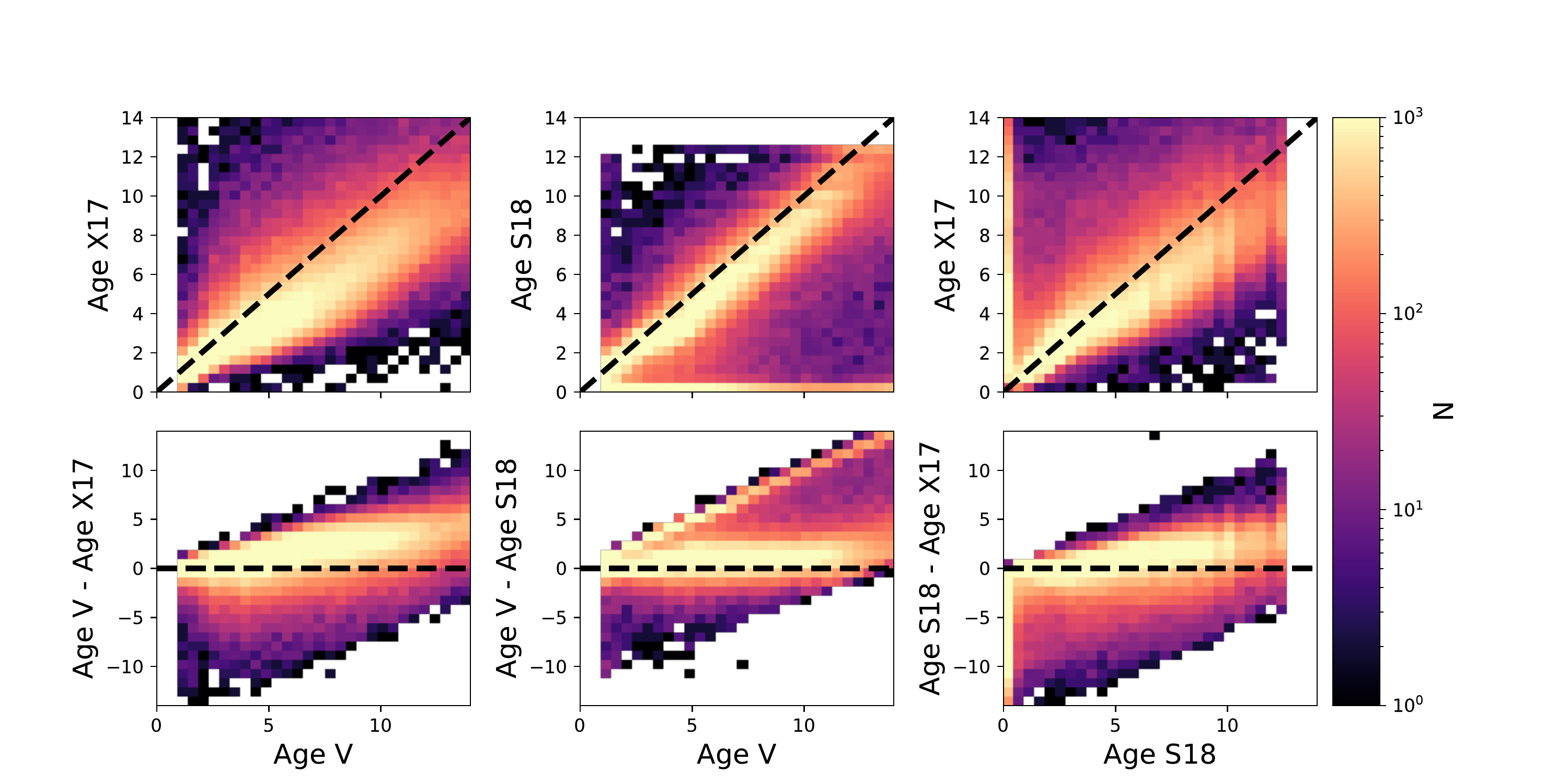}
	\caption{
          Comparison of the ages used in this work (``V'') with ages determined in \citet{xia2017} (``X17'') and \citet{san2018} (``S18'') for overlapping stars. Overall we see qualitative agreement, with expected differences arising from the use of different isochrones (we have used the Dartmouth models, \citealt{xia2017} used Yonsei-Yale models, \citealt{san2018} used PARSEC models), and different procedures. Errors are incorporated in this plot.
        }
	\label{fig:age_comp}
\end{figure*}

In Figure \ref{fig:age_comp} we plot the age versus age and age difference versus age plots for the data in these age catalogs, and we see general agreement between all three. Our catalog agrees best with the ages derived by \citet{san2018}, with a fairly constant offset of about 1.8 Gyr (where our ages are older). There is a more significant and non-uniform difference between our derived ages and those of \citet{xia2017}, which may in part arise from those authors' unique approach to calibrating the stellar temperature values (to the scale of \citealt{hua2015}; see their paper for more details), as well as their usage of different isochrones in their age calculation.

This comparison is particularly interesting as these three works used three different sets of isochrones. As mentioned, we use the Dartmouth family of isochrones. \citet{xia2017} utilize the Yonsei-Yale isochrones (\citealt{yi2001}, \citealt{kim2002}, \citealt{yi2003}, \citealt{dem2004}), and \citet{san2018} make use of the PARSEC isochrones \citep{bre2012}.

Various differences exist between these families of stellar models. $\alpha$ abundances, for example, are available for Dartmouth and Yonsei-Yale isochrones, but are not yet readily available for Padova isochrones. Young stellar ages (younger than 1Gyr) are available for Yonsei-Yale and PARSEC isochrones, but not for Dartmouth models (although, the MIST isochrones from the lead author of the Dartmouth isochrones should be adding $\alpha$ element abundances soon). Dartmouth models are also generally found to ``run hot,'' leading ages derived with Dartmouth isochrones to usually be slightly older than those derived with other models.

While a more detailed comparison is beyond the scope of this work, it is encouraging to see that these three catalogs produce similar age determinations, and it is interesting to see the magnitude of difference that can be expected from using different procedures and models in the determination of stellar age.

Moving beyond comparison with other age catalogues derived using isochrones, we may also compare our derived ages with those calculated using astroseismology. Astroseismic ages (ages calculated based on the pulsations of stars) are considered to be some of the most accurate age estimates possible with current data. The downside of astroseismic ages is that high-quality, time-series photometry is required, which limits the sample size considerably compared to other methods.

\begin{figure}
	\includegraphics[width=\linewidth]{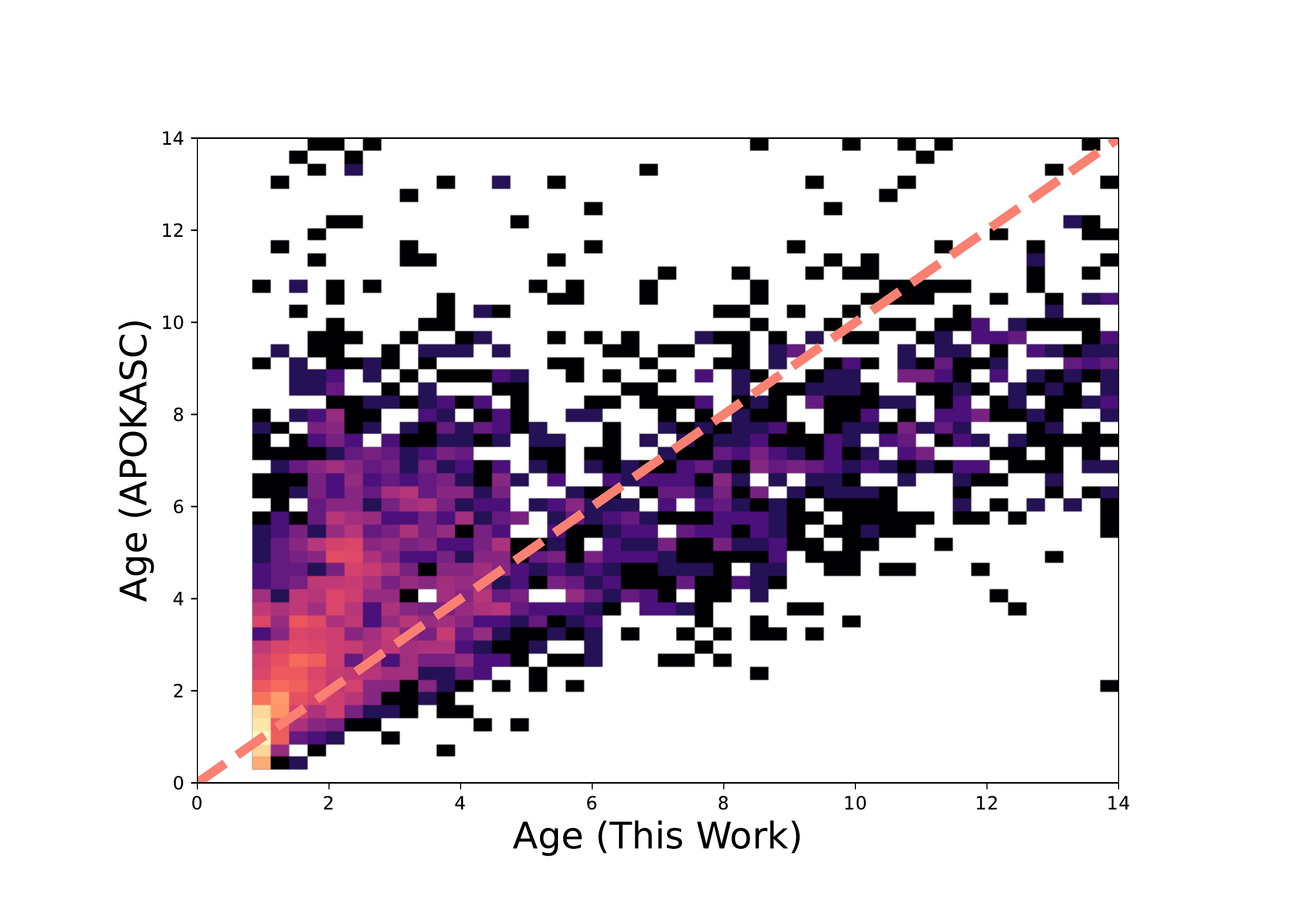}
	\caption{
          Comparison of ages in this work and the astroseismic ages from the APOGEE-Kepler overlapping sample (the APOKASC catalogue). The APOKASC sample is a subset of APOGEE data, which means it consists mainly of red-giant stars. Red giant stars are expected to be problematic to calculate ages for using isochrone techniques. Indeed, we see a fair amount of disagreement between the ages derived with these two techniques, particularly for giants assigned young ages by our pipeline.
        }
	\label{fig:apokasc}
\end{figure}

Luckily, LAMOST has observed a large number of objects in the Kepler field, which is where the majority of astroseismic age-work has been done. In Figure \ref{fig:apokasc} we compare our derived ages with the astroseismic ages available in the second APOGEE-Kepler astroseismic catalogue, APOKASC \citep{pin2018}. The APOKASC catalogue is mostly giant stars.

We see reasonable qualitative agreement between the two catalogues at intermediate and older ages (as determined in this work), and a fair amount of disagreement among the giant stars that we calculate to be young. The disagreement is to be expected since isochrone age determinations are notoriously less reliable in areas where isochrones of different ages overlap, such as the giant branch or main sequence, and the APOKASC catalog contains mainly red giants.

While there is little to be done regarding the young-age disagreement with our current dataset, we note that only 15\% of our data are giants (having surface gravities less than 3). We also rerun the main analysis of this work with a subsample of data which avoids the giant branch and main sequence, in Section \ref{sec:reliable}, and find our results with both the full sample and high-quality subsample to be similar.

\begin{figure}
	\includegraphics[width=\linewidth]{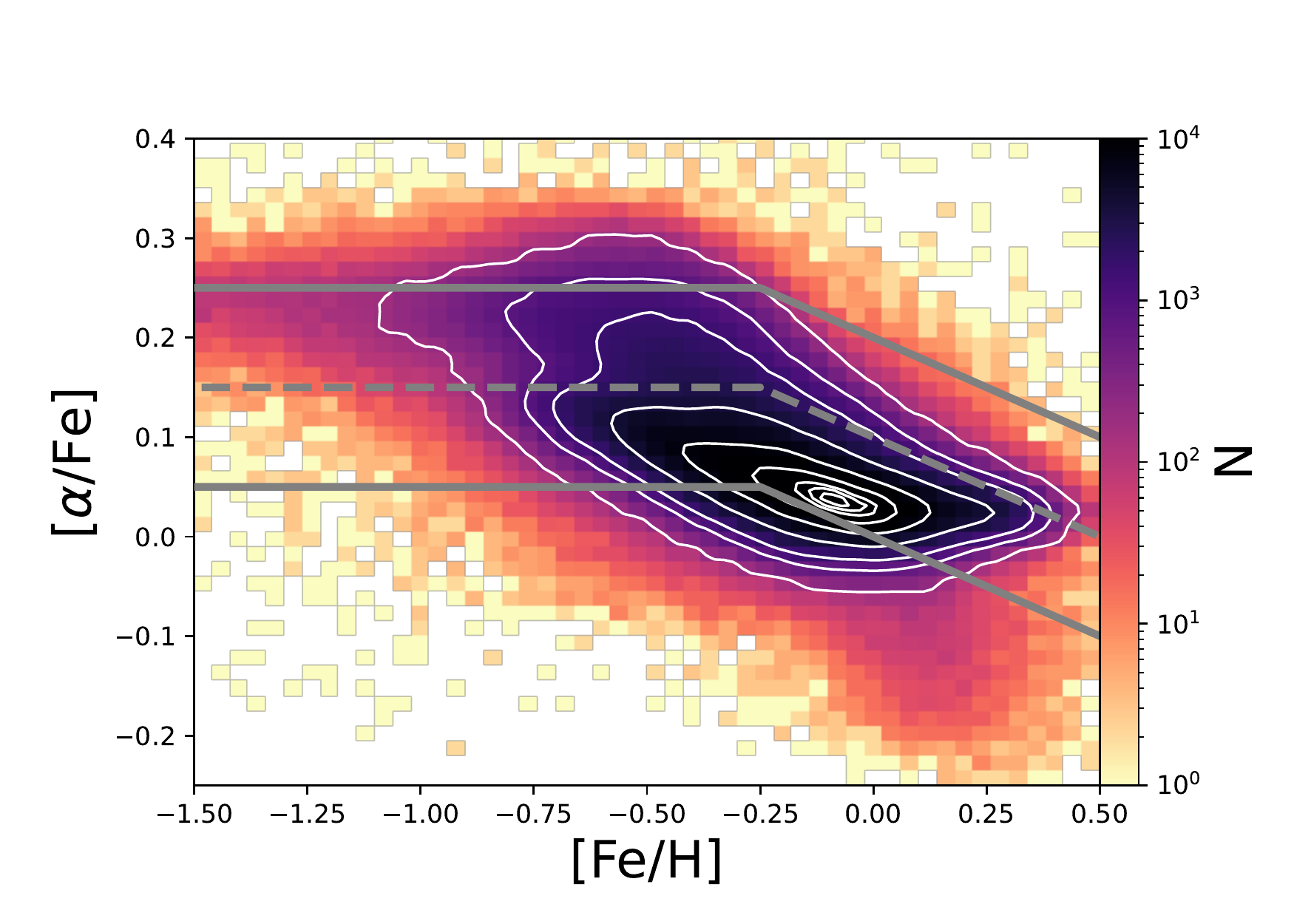}
	\caption{
          The [$\alpha$/Fe] vs [Fe/H] plot for our dataset. A strong concentration of stars with a downward slope, visible in a saturated black, is the ``chemical thin disk.'' Above this concentration are the more $\alpha$-enhanced ``chemical thick disk'' stars. We have overlaid three lines which split these data into four groupings in chemical space. The primary (dashed) line is selected by eye with the points ([$\alpha$/Fe], [Fe/H]) = (-0.5, 0.15), (-0.25, 0.15), (0.25, 0.05). The lines above and below this dividing line are shifts of $\pm$ 0.1 in [$\alpha$/Fe]. The contours indicate 5\%, 25\%, 50\%, 75\% and 95\% of the maximum density.
        }
	\label{fig:bifurcation}
\end{figure}

\subsubsection{Orbits and Coordinates}

We use a right-handed coordinate system with the Sun located along the x-axis at $(X, Y, Z) = (-8.27, 0, 0)$ kpc. The solar system rotates with a velocity of $(V_{R}, V_{\phi}, V_{Z}) = (0, -236, 0)$ km s$^{-1}$ (alternatively, $(U, V, W) = (0, 236, 0)$ km s$^{-1}$) and it has a motion relative to the local standard of rest of $(U, V, W) = (13.0, 12.24, 7.24)$ km s$^{-1}$. These values are adopted from \citet{sch2017}.

We assign cartesian coordinates for our stars using the on-sky coordinates and distances from the catalog of \citet{bai2020}. Cartesian velocities are calculated following the procedure of of \citet{joh1987}. From these cartesian velocities, we calculate the spherical and cylindrical velocities in the usual way.

To account for measurement errors, we create one hundred mock observations per star with the relevant parameters pulled from their error distributions and accounting for the correlations between the measurements (as reported in the Gaia astrometric catalog). The final phase-space coordinates and errors are the mean and standard deviation of these one hundred Monte-Carlo error expanded mock observations.

We calculate the orbits based on these final phase-space coordinates with the python package $galpy$ (described in \citealt{bov2015}\footnote{https://github.com/jobovy/galpy}). Within this package, we use the provided potential MWPotential2014 which is a superposition of: an exponentially cut off power law bulge \citep{mcm2011}, a Miyamoto-Nagai disk \citep{miy1975}, and an NFW halo \citep{nav1997}.  The potential is modified to reflect our coordinate system values for solar position and velocity, noted above. We will also make use of the potentials of \citet{irr2013} and \citet{mcm2017}, which are available in the same code library and are similarly modified for our solar values.

For this paper, we will be measuring the iron abundance [Fe/H] as a function of the guiding radius, \rguide, defined as the radius of a circular orbit with the same angular momentum as the observation within the same Galactic potential. This is slightly more precise than approximating \rguide \ as the average of the orbital apocenter and pericenter radii, and accounts for the effect whereby stars are generally closer to pericenter than apocenter along their orbits. This \rguide-based gradient is less affected by blurring (epicyclic motions) than an \rpres \ based gradient, and its use can help reveal the intrinsic underlying gradient.

\subsection{Quality Cuts}

Having obtained chemistry, phase information, orbits, and age information for our data, we perform some quality cuts to prepare our final data set.

We remove duplicate observations within 5 arcseconds, preserving the observations with the highest signal to noise ratio (defined as the quadrature sum of the signal to noise ratio in $g$, $r$, and $i$).

From our initial sample of $\sim$4 million stars with orbits and ages (the crossmatch of LAMOST and Gaia with full spectroscopic and astrometric information), we select only stars with signal to noise ratios greater than 50 in all three individual photometric passbands and with age errors less than 25\%. This leaves a sample of 1,328,841 stars.

\section{Results}\label{sec:results}

To study how the metallicity gradient evolves in different populations in the Milky Way, we investigate different selections for thin and thick disk populations and later investigate the effects of varying those selections. The thin disk is often separated from the thick disk in three distinct spaces:
\begin{itemize}
\item Chemistry - the [$\alpha$/Fe]-[Fe/H] plot (Figure \ref{fig:bifurcation}) of local stars shows a diffuse cloud of an $\alpha$-rich population (commonly referred to as the chemical thick disk) and a strong concentration of an $\alpha$-poor population (commonly known as the chemical thin disk). This abundance grouping of stars is well documented in local observations (\citealt{hay2015} or \citealt{vin2021}, for example). $\alpha$ elements are produced most prodigiously in type-II supernovae, which occur in high mass stars over relatively short timescales and initially seed the interstellar medium with $\alpha$ elements. Over a longer timescale, type-Ia supernovae, which require a star to evolve to the white dwarf stage in a binary system, begin to inject large amounts of iron into the interstellar medium, reducing the observed [$\alpha$/Fe] ratio. The chemical thick disk, with relatively higher $\alpha$ element abundances, is therefore thought to have been formed in the early epochs of Milky Way formation before the onset of type-Ia supernovae, while the thin disk formed after.
    
\item Vertical Height (\zmax) - stellar density studies have shown that star counts vertically out of the plane are better fit to two overlapping exponential profiles than to one (\citealt{yos1982}, \citealt{gil1983}). Similar studies in star counts have led to the understanding that observations may be well fit by a population whose orbits inhabit a longer scale length and a larger scale height (the physical thick disk) and another overlapping population whose orbits fill a shorter scale length and thinner scale height (\citealt{jur2008}; note that the chemical thick disk has a larger scale height and \emph{shorter} scale length than the chemical thin disk).

\item Kinematics - the thin disk is known to rotate at speeds up to the circular rotation speed of the Galactic potential, with an extended distribution toward slower speeds accompanied by increasing vertical and radial random motions (\citealt{oor1922}, \citealt{str1924}). It has been noted that while both disks inhabit a range of velocities, their velocity distributions are distinct (\citealt{wys1986}, \citealt{ojh1996}) and they may be separated on that basis.

\end{itemize}

Of these three, we found that chemical separation and orbital separation (vertical height, determined by the stars' maximum $Z$ coordinate along their orbits) were the most effective. Kinematic separation on the basis of the ``Toomre'' diagram\footnote{The ``Toomre Energy Diagram'' seeks to separate the kinematic populations of the galaxy in a plot which compares the rotational velocity of a star to the non-rotational velocity of a star (represented by the radial and vertical velocities added in quadrature, see \citealt{nis2004} for example).} was attempted, but it suffered significant cross-contamination and wasn't very useful for our purposes here. This could be because of the relatively large radial extent of our data causing kinematic separations which are effective at the solar neighborhood to be less useful for data extending over a large range of radii (over which the velocity distributions of the thin and thick disks may differ). We attempted to correct for this by replacing $V_{\phi}$ with $V_{\phi}-V_{circ.}$ where $V_{circ.}$ is calculated for each star based on their radius, but this correction did not seem to alleviate the cross-contamination, possibly because asymmetric drift effects make this type of kinematic separation more complex.

We select ``pure'' thin disk and ``pure'' thick disk samples using these two (chemical and physical) criteria, and show how the radial metallicity gradient changes for those stars with age. We find that the gradient for the thin disk gradually grows shallower with age at a rate of 0.003 dex kpc$^{-1}$ Gyr$^{-1}$ (Section \ref{ssec:mgrad_thin}). For a thick disk selection we find a positive gradient which is similar (near 0.02 dex kpc$^{-1}$) for all ages (Section \ref{ssec:mgrad_thick}).

We follow up by dissecting the effects of different aspects of our selection by slicing through various chemistry and \zmax \ cuts; we check the influence of our fitting parameters by investigating various potentials, solar radii, and solar rotation speeds; and we briefly note how our findings change with more restrictive age quality cuts and less precise radii determinations.

\subsection{Evolution of the Metallicity Gradient in the Thin Disk}\label{ssec:mgrad_thin}

\begin{figure*}
	\includegraphics[width=\linewidth]{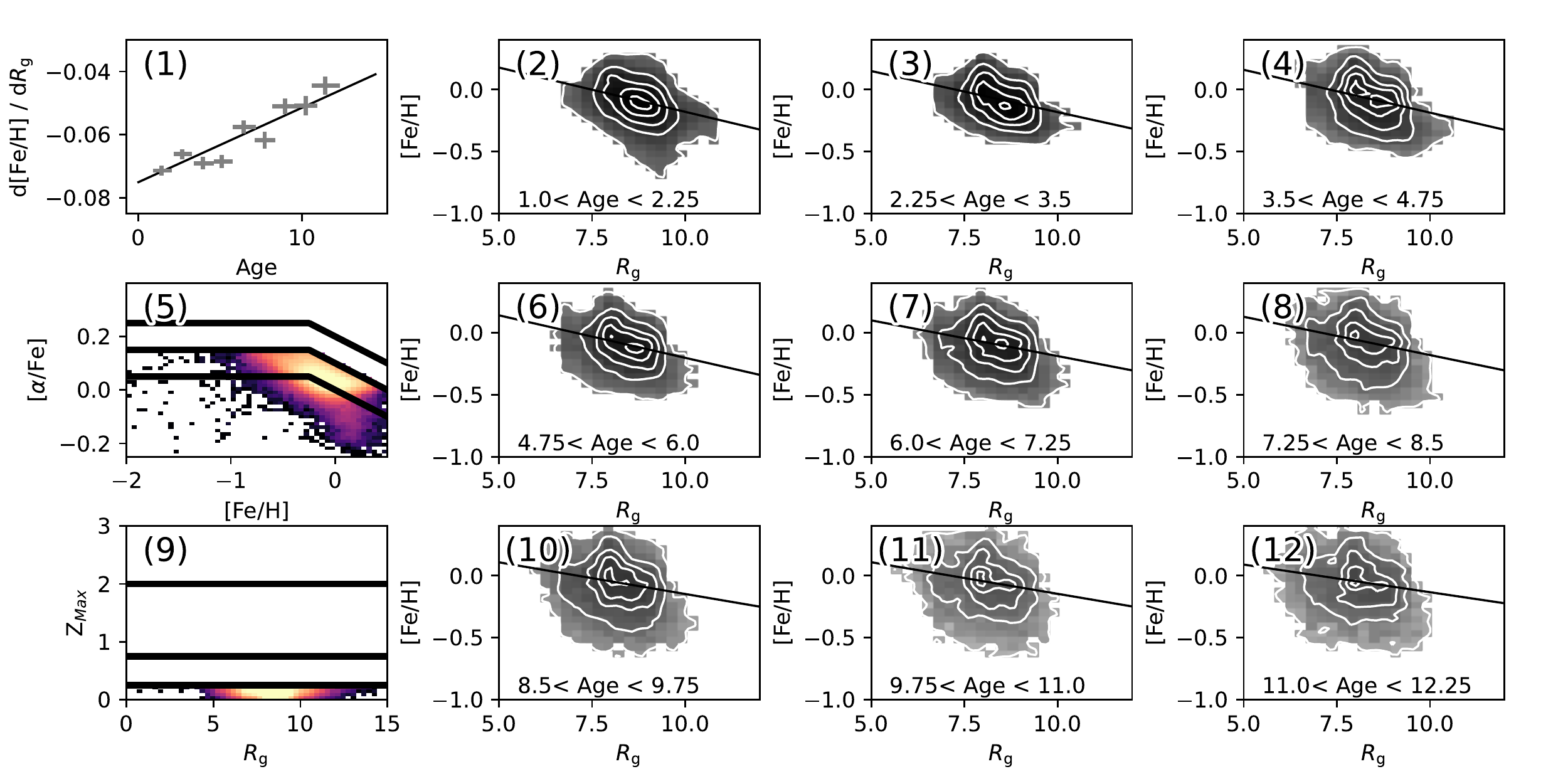}
	\caption{
          The selections for our thin disk ([$\alpha$/Fe] $<$ 0.15, [$\alpha$/Fe] $<$ -0.2$\cdot$[Fe/H] $+$ 0.1, \zmax \ $<$ 250 pc) and the resulting metallicity gradient (\emph{left top}). \emph{Left-middle, left-bottom}: The distribution of our thin disk in chemical and orbital spaces, respectively. We find that, with this chemical-orbital selection, our observed metallicity gradient is monotonic and linear with age. \emph{Right panels} show the [Fe/H] vs \rguide \ distributions for various age bins. The youngest stars show a strong negative gradient, while the older stars show a more diffuse and less ordered distribution.
        }
	\label{fig:grad_thindisk}
\end{figure*}

Our ``pure thin disk'' sample comprises a selection of $\alpha$-poor stars ([$\alpha$/Fe] $<$ 0.15 and [$\alpha$/Fe] $<$ -0.2$\times$[Fe/H] $+$ 0.1), confined to having orbital \zmax \ less than 250 pc from the plane. These thin disk data are then binned in 1.25 Gyr age bins from 1 to 12.25 Gyr and each bin is fit in [Fe/H] as a function of \rguide \  using orthogonal distance regression\footnote{https://docs.scipy.org/doc/scipy/reference/odr.html} (a linear fit with observational errors in both x and y directions considered; \citealt{bog1990}).

We note that similar works, such as those by \citet{yu2012} and \citet{xia2015}, opted to bin their data in radius, and then fit the bins. When we follow this procedure, we find results similar to those authors. In this work, however, we have decided to utilize our current method in an effort to preserve as much information from the data as possible. One downside to our approach is that the highly populated areas of our data (for example, the data near the Sun), tend to control the fits more than the sparsely populated regions (such as the data near the outskirts of the Galaxy).

The gradients resulting from this selection and fit are shown in Figure \ref{fig:grad_thindisk}, along with the stars' distribution in chemistry space ([$\alpha$/Fe] vs [Fe/H]) and orbital space (\rguide \ vs \zmax), as well as the distribution of each bin's members in [Fe/H] vs \rguide.

It is somewhat surprising that the data in each age bin seem to have similar extents in [Fe/H], that is, there are young stars with low metallicities, and old stars with high metallicities. While exotic stars, such as ``young'' high-$\alpha$ stars (which will be discussed further in the next section) can explain some of this, it is unexpected that they would contribute a significant amount.

It is plausible that, due to net outward migration over time in the Galaxy, our oldest stars are from more interior regions, and so could naturally have higher metallicities than younger stars born farther out in the Galaxy. Another reason for the large extents in metallicity may be observational errors and possible underestimation of those errors.

One concern may be that there are some systematic effects regarding the atmospheric parameter determination, such as a degeneracy with log(g) or T$_{eff.}$. A brief inspection of the data in Figure \ref{fig:grad_thindisk} reveals that each panel (2, 3, 4, 6, 7, 8, 10, 11, 12) has a very slight gradient from high surface gravity at the center of the distribution to lower surface gravity at the edges (in \rguide). Such a trend could be an effect of the higher surface brightness of lower surface gravity stars allowing us to see them to greater distances. The weak nature of that gradient may be because we use \rguide \ instead of \rpres. There seems to be no gradient in [Fe/H] with surface gravity though (except in the youngest age bin, where the most metal poor stars appear as a clump of higher surface gravity dwarfs than the rest of the stars), and no large difference in the surface gravity of stars between panels.

Besides the large extents of the data in metallicity, the trends seem as expected: first, the data seem to move more metal poor as age increases (looking, for example, at the largest contour-line's lowest extent in each panel), and second the data grow more diffuse in \rguide \ with increasing age as is expected (if we observe the same volume for both old and young stars, the older ones should have hotter orbits and so fill a larger extent in \rguide).

The x-error bars in the top left panel indicate the 68\% confidence interval of the ages of stars in that bin. The stars are first partitioned into different age bins based on their maximum likelihood ages; then, in each bin, the ages are expanded about their errors and the 68\% confidence intervals of these error-expanded, in-bin ages are plotted as the error bars (Figure \ref{fig:grad_thickdisk} shows this more clearly, with the oldest bins having noticeably larger x-errorbars). The y-error bars in the same panel indicate the 1$\sigma$ range of slopes as determined by the orthogonal distance regression fitting.

We see that the pure thin disk selection results in a negative metallicity gradient which is steepest for the youngest stars, around -0.075 dex kpc$^{-1}$. This value is in agreement with recent works such as \citet{che2020} and \citet{zha2021}, who find gradients of -0.07 and -0.074 dex kpc$^{-1}$ (respectively) for clusters younger than 500 Myr. This gradient flattens toward older age stars with a slope of 0.003 dex kpc$^{-1}$ Gyr$^{-1}$. There could be some hints of flattening of this trend in the intermediate age bins around 4.75 and 6 Gyr, but overall the trend is fairly monotonic.

This relation is the inverse of the birth expectations from inside-out growth, with the oldest stars being formed with the steepest radial metallicity gradients and the youngest being born with the shallowest, and so represents probable observational evidence of stellar migration in action.

Similar work by \citet{xia2015} and \citet{wan2019} (both of which used LAMOST data) found their intermediate age stars to have the steepest radial metallicity gradients, with younger and older stars having shallower gradients. A major difference between those works and ours is that they used \rpres \ while we prefer \rguide. When we recalculate the results shown in Figure \ref{fig:grad_thindisk} using \rpres, we similarly find a ``U'' shaped distribution of gradients (with slopes from young to intermediate age to old of -0.08 to -0.11 to -0.08), with the intermediate stars having the steepest metallicity gradients. We suspect that this could be an artifact of the complex selection function of our spectroscopic data, which is inherently non-uniform in cartesian space.

In CoRoT and APOGEE, \citet{and2017} studied the metallicity gradient in \rpres \ of giants and saw a metallicity gradient which generally flattened with increasing age, interrupted by slight dip toward a steeper gradient for intermediate age stars (from -0.058 to -0.066 to -0.03 dex kpc$^{-1}$ for young to intermediate to old stars). A similar flattening trend interrupted by a small dip at intermediate ages has been seen in Geneva Copenhagen Survey data when using \rguide \ (\citealt{nor2004} find slopes from young to intermediate age to old of -0.076 to -0.099 to 0.028 dex kpc$^{-1}$; \citealt{cas2011} find similar results, see their Figure 18). \citet{mac2017} also found evidence for a shallowing metallicity gradient with increasing age in their low-$\alpha$ sample, as well as evidence for older populations to have larger Galactocentric radial dispersions.

\subsection{Evolution of the Metallicity Gradient in the Thick Disk}\label{ssec:mgrad_thick}

\begin{figure*}
	\includegraphics[width=\linewidth]{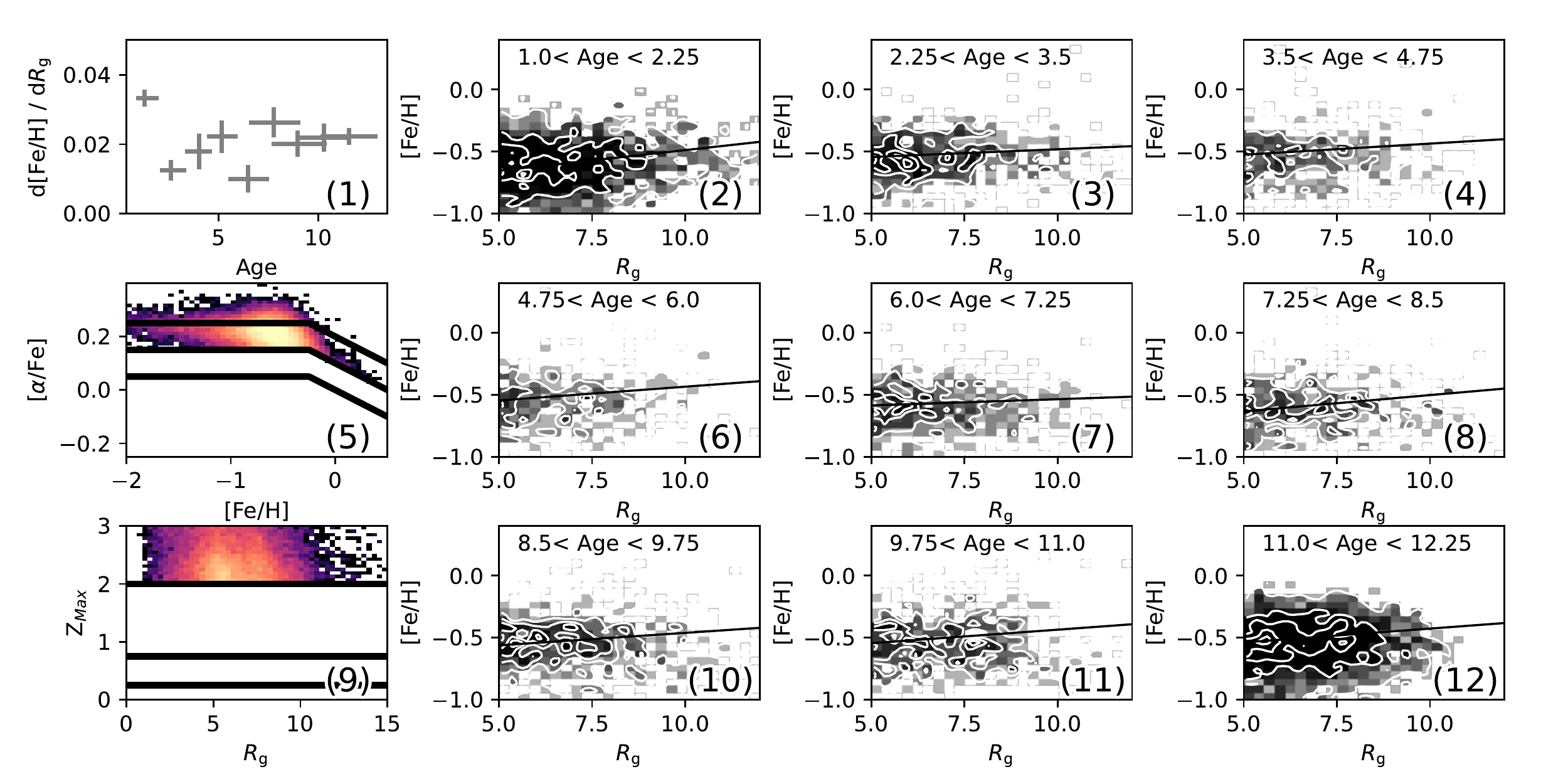}
	\caption{
          The selections for our thick disk ([$\alpha$/Fe] $>$ 0.15 or [$\alpha$/Fe] $>$ -0.2$\cdot$[Fe/H] $+$ 0.1, \zmax \ $>$ 2 kpc) and the resulting metallicity gradient (similar to Figure \ref{fig:grad_thindisk}). The population appears well mixed in age, with all populations having similar metallicity gradients. Interestingly, these gradients are always positive, which would seem to imply that the population is not well mixed chemically in radius.
        }
	\label{fig:grad_thickdisk}
\end{figure*}

To select the thick disk, we merely reverse our thin-disk criteria, collecting $\alpha$-enhanced stars who's orbits extend more than 2 kpc from the plane. We calculate the metallicity gradients in a manner similar to that used in Figure \ref{fig:grad_thindisk} and plot the results in Figure \ref{fig:grad_thickdisk}.

The figure shows that our thick disk sample is generally more metal poor than the thin disk sample, which is expected when selecting a high [$\alpha$/Fe] population. The thick disk also appears to be more centrally concentrated and extends outside the plot axes (which have been set to match the extent in Figure \ref{fig:grad_thindisk}) toward the Galactic center -- this more central concentration is probably related to the shorter scale length generally found for the chemical thick disk relative to the chemical thin disk.

We note that the youngest age bin (panel 2) seems to have an overabundance of low surface gravity giants. Since our age estimates for giants are not as reliable as for other types of stars, this specific bin might be populated with stars which belong in other panels of this Figure.

We find that the weak radial metallicity gradient for thick disk stars is similar for all ages (near 0.02 dex kpc$^{-1}$). This is expected, since we presume the thick disk to be well mixed. The positive nature of the gradient at all ages, however, is surprising.

We do seem to see some indication of inside-out formation, in that the older stars are more inwardly concentrated than the younger stars.

\begin{figure}
	\includegraphics[width=\linewidth]{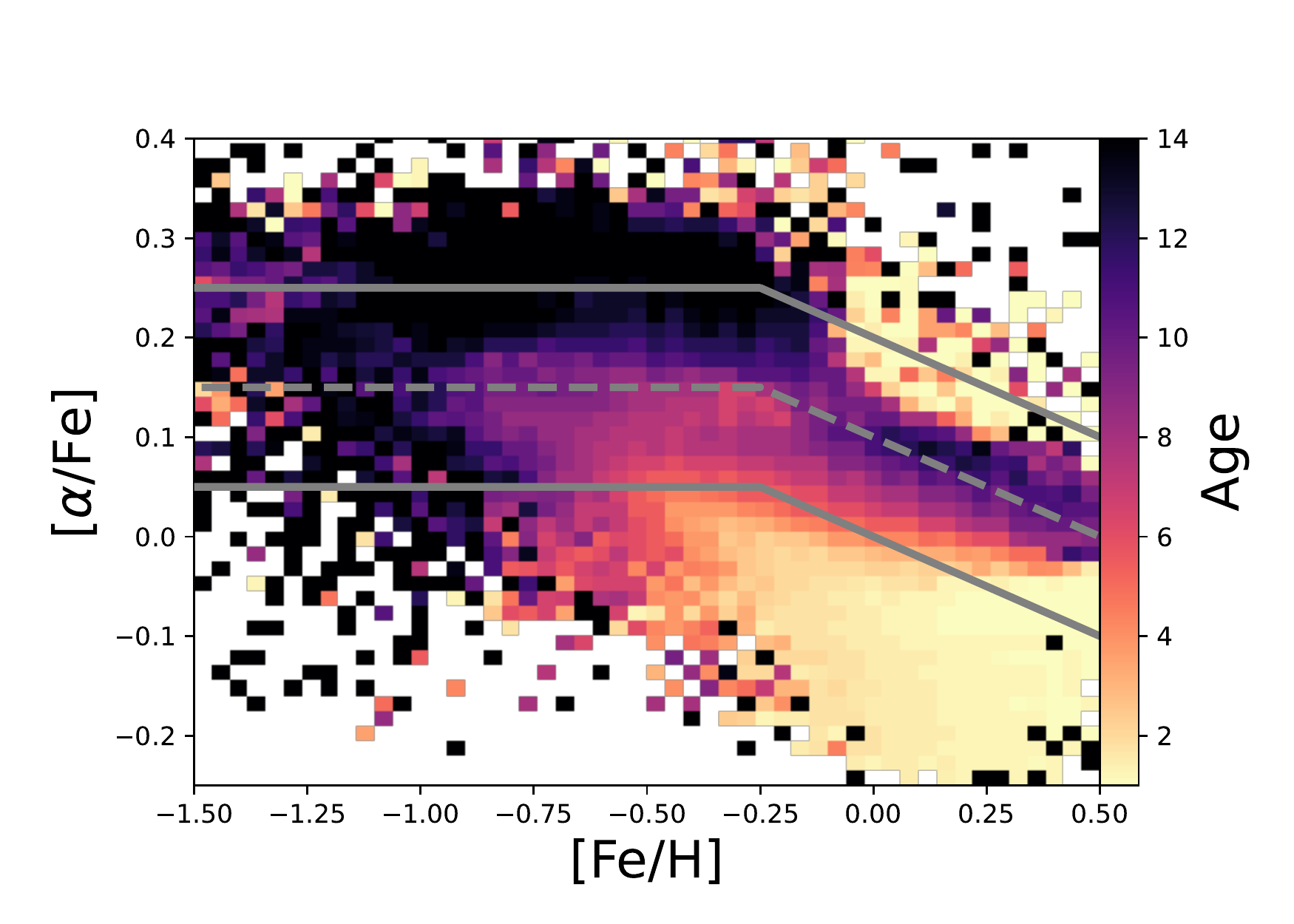}
	\caption{
          The chemistry distribution of our data, color-coded by median age. While we see the expected old-age-high-$\alpha$ and young-age-low-$\alpha$ clusters, we also see unexpected patterns, such as a population of young, metal-rich, high-$\alpha$ stars, which will fall into our chemical thick disk selection.
        }
	\label{fig:bifur_age}
\end{figure}

In Figure \ref{fig:bifur_age}, we plot the chemistry space of our data, and color code it by the median age in each bin. This figure reassuringly shows a concentration of old stars in the high-$\alpha$ section, and a concentration of young stars with lower $\alpha$ abundances. However, the high [Fe/H] edge of the high-$\alpha$ section (upper right) shows an overabundance of young stars, a puzzling population, perhaps related to the young high-$\alpha$ stars noted by \citet{chi2015} in CoRoT and \citet{mar2015} in Kepler. We have tested removing this young population and replotting Figure \ref{fig:grad_thickdisk} with the additional cut of [Fe/H] $<$ -0.1, but the results were unchanged.

It is worth mentioning that there is a strong error-signal, when calculating ages, where atmospheric parameters are naturally scattered by their errors outside of the parameter space covered by isochrone families. This leads to large pile-ups of stars in the oldest and youngest bins, as the youngest and oldest isochrones define the edge of the parameter space covered by isochrones. Our results, however, are unchanged by excluding these bins, and the bins up to 5 Gyr of age are populated by the chemical thick disk, far away from such edge effects.

The origins of young, high-$\alpha$ stars are still debated: \citet{chi2015} postulated that these stars could arise from inert gas near the ends of the Galactic bar; other authors note that they are frequently found in binary systems \citep{jof2016} and show evidence of mass transfer (\citealt{yon2016}; see also \citealt{hek2019}) which suggests that these stars are truly Blue Straggler stars \citep{mat2018, sun2020}. This could lead to an observationally young population of stars which is kinematically old and similar in chemistry to the older thick disk.

While we think these stars are reasonable to leave in our data, the reader should be aware of this odd population, and consider its implications with respect to our findings.

\section{Discussion}\label{sec:discussion}

\subsection{Effect of Various Selectors and Galactic Parameters}

\subsubsection{Maximal Vertical Height}\label{sec:zmax}

\begin{figure*}[!htb]
	\includegraphics[width=\linewidth]{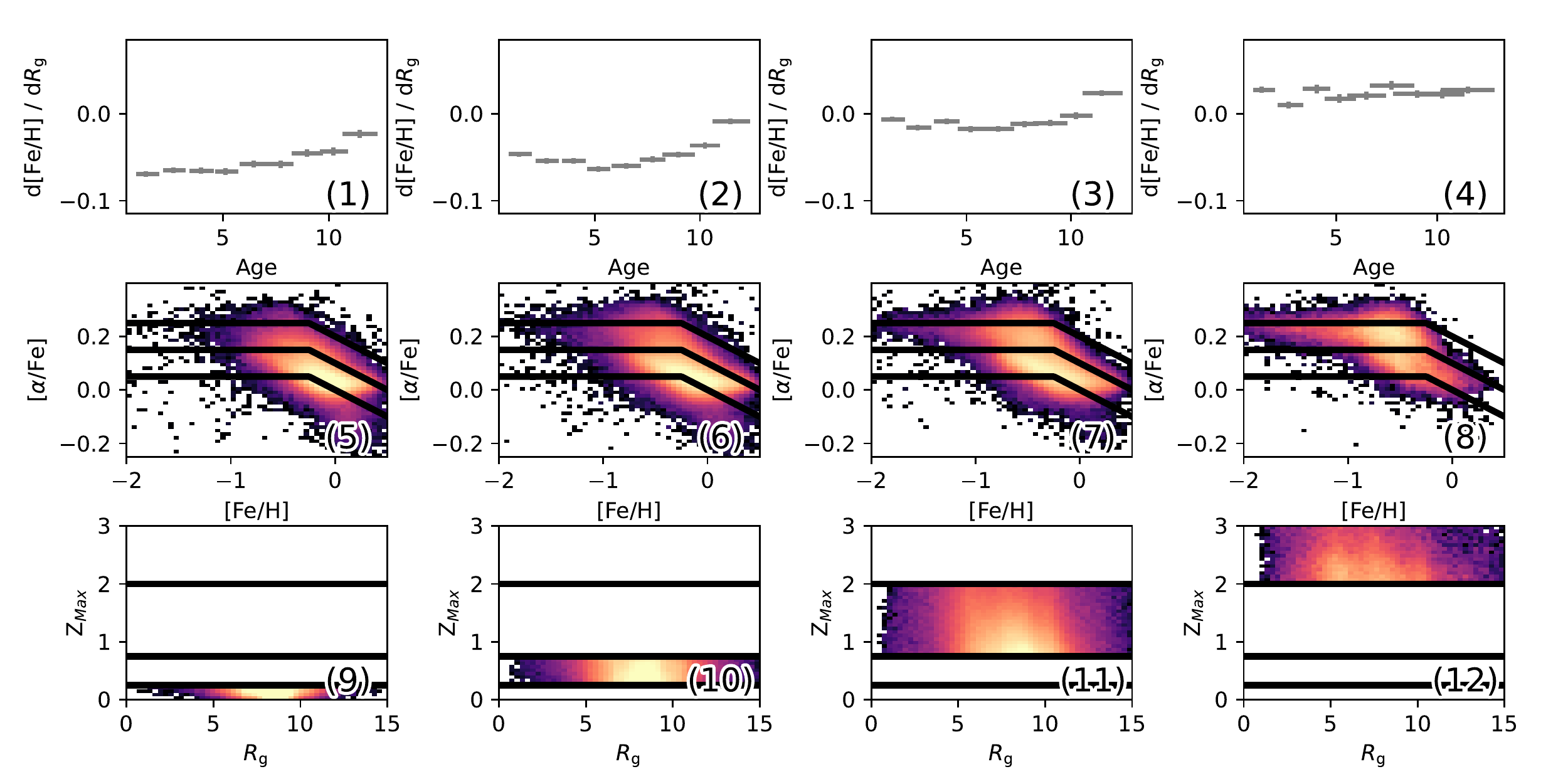}
	\caption{
          \emph{Top:} The metallicity gradient of various subsets selected by \zmax \ from orbitally thinner (``cold'') at the \emph{left} to orbitally thicker (``hot'') at the \emph{right}. The \emph{middle} and \emph{bottom} rows show the chemical and orbital (respectively) distribution of stars in their respective columns.
        }
	\label{fig:grad_zmax}
\end{figure*}

In our thin disk selection, we included a stringent cut on \zmax \ to select stars whose orbits are confined to within 250 pc from the plane. As we have mentioned, the mechanics of churning may favor stars on colder orbits, specifically stars with low vertical velocity dispersions \citep{ver2014} or low eccentricities, therefore we suspect that \zmax \ should be strongly correlated with observational evidence of that. That is, we expect the stars on the vertically coldest orbits to show the strongest evidence for the [Fe/H] gradient flattening relative to their gradients at birth. This flattening over time could even cause us to observe older stars with shallower metallicity gradients than younger stars, despite presumably having been born with steeper radial metallicity gradients (as the gradient in the ISM is commonly believed to flatten out over time).

That isn't to say that migration doesn't affect stars on hotter orbits, such as thick disk stars, as was shown in \citet{sol2012}, merely that stars on colder orbits may be preferred.

In Figure \ref{fig:grad_zmax}, we slice our data into four subsets on the basis of \zmax, find the corresponding metallicity gradients, and also plot their chemical and kinematic distributions, as in Figure \ref{fig:grad_thindisk}. As we expected, the tighter \zmax \ constraints accentuate the trend for older stars to have flatter metallicity gradients.

Two further trends become apparent here. First, we note that the chemical and orbital spaces are interlinked: the low \zmax \ selected data in Figure \ref{fig:grad_zmax} are also more $\alpha$-poor than the rest of the sample, while the high \zmax \ data are more $\alpha$-rich. Secondly we note that as we transition from the ``thin disk'' sample to the opposite ``thick disk'' sample, the gradient flattens out to become more similar at all ages and increases first toward zero, and then to positive values.

For intermediate \zmax \ stars, we see a slight ``uptick'' with the youngest stars having a shallower [Fe/H] vs $R$ gradient than intermediate age stars. We believe this could be contamination from the thick disk (since the gradient is more positive than the most ``thin disk'' selection). While thick disk stars are not expected to be prominent in this age range, they could be ``observationally young'' yet ``kinematically old'' if they are indeed Blue Stragglers, as proposed by some authors and discussed above.

Note that a \zmax \ cut, particularly a strict one, could remove some of the strongest migrators in the data if they also flare as they migrate. This would mean that, for example, a high metallicity star which migrates from the center to the outskirts, which should raise the [Fe/H] value in that outer bin, may be removed. This would work to reduce the metallicity in outer bins and increase the steepness of our gradients. Figures \ref{fig:grad_zmax} and \ref{fig:grad_chem} can be investigated to consider the magnitude of this effect.
  
\subsubsection{Chemistry}\label{sec:chemistry}

\begin{figure*}[!htb]
	\includegraphics[width=\linewidth]{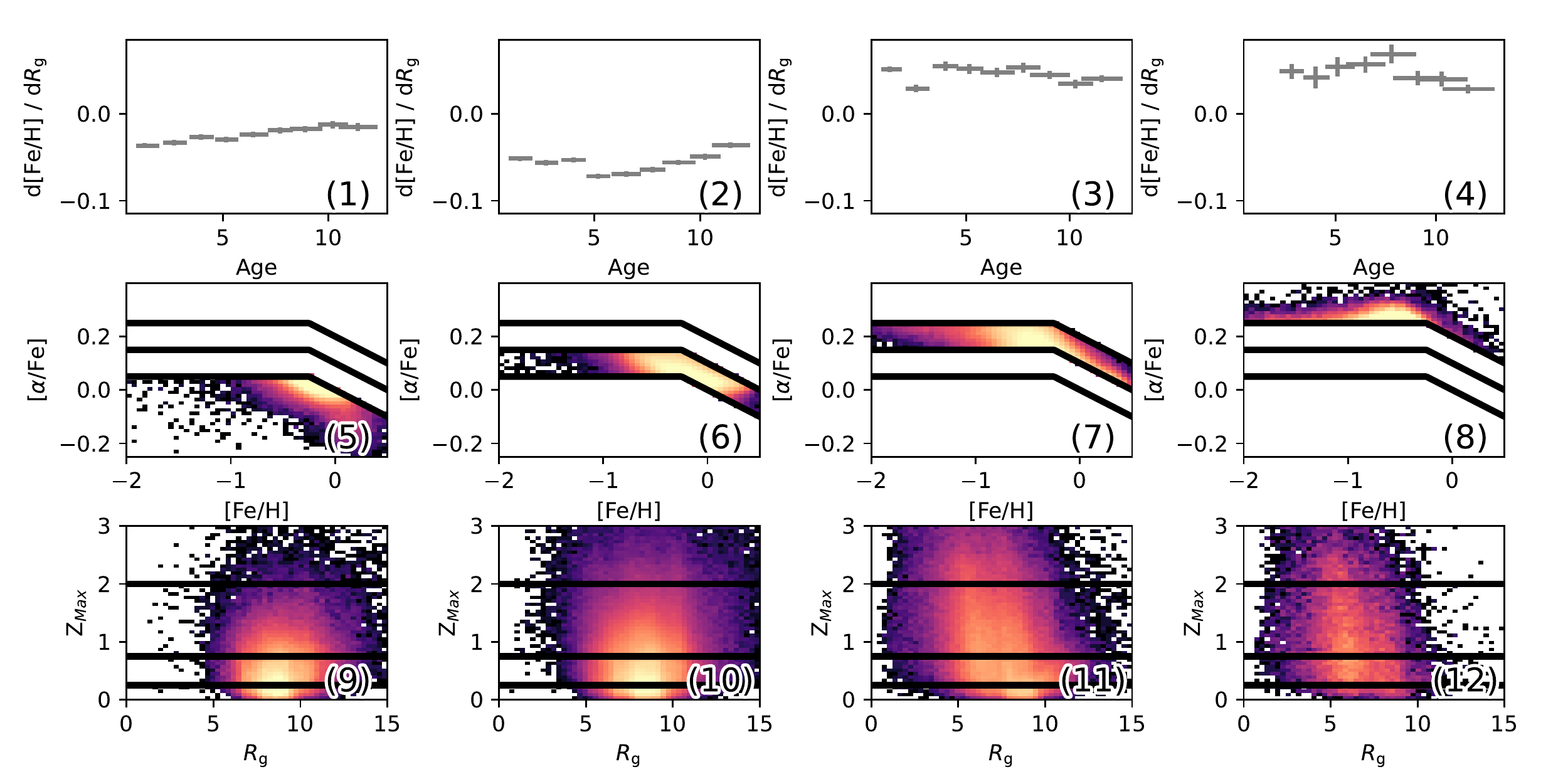}
	\caption{
          Similar to Figure \ref{fig:grad_zmax} only with the data split based on chemistry instead of orbital \zmax.
        }
	\label{fig:grad_chem}
\end{figure*}

Two populations may be seen in a bifurcation of the [$\alpha$/Fe]-[Fe/H] space in the local neighborhood. In Figure \ref{fig:bifurcation} we showed the abundance plot for our data and noted a dense concentration of stars at low $\alpha$ abundances and a sparse cloud at high $\alpha$ abundances. We split these data into four sectional divisions. The strong concentration of stars with $\alpha$ abundances below 0.15, would be the conventional ``chemical thin disk.'' Above this, at iron abundances below zero, would be the ``chemical thick disk.''

On the basis of the divisions shown in Figure \ref{fig:bifurcation}, we again slice our data in four subsets and then fit the metallicity as a function of \rguide \ in age bins in Figure \ref{fig:grad_chem}.

We see that the $\alpha$-poor stars inhabit the thin disk like areas in the orbital characteristic plots, with lower \zmax. They also have the strongest trend in [Fe/H] vs \rguide \ as a function of age, similar to the orbitally selected thin disk. As we progress toward thick disk chemistries ($\alpha$-rich stars), this gradient again moves toward positive values and flattens out.

For intermediate $\alpha$-abundance stars, we again see an ``uptick'' where the youngest stars have a shallower gradient than intermediate age stars. 

An oddity in Figure \ref{fig:grad_chem} provides a nice opportunity to investigate how all of these figures fit together. One may note that, in the second column of Figure \ref{fig:grad_chem}, the calculated gradients (panel 2) appear steeper for all ages of stars than the gradients shown in Figure \ref{fig:grad_thindisk} (panel 1), except the youngest bins which are slightly shallower. Since the sample in this column is quite similar to the sample used in Figure \ref{fig:grad_thindisk}, we wonder if there is a disagreement.

If we look to the first column in Figure \ref{fig:grad_chem} (panel 1), which is the second half of the chemical space covered by the sample in Figure \ref{fig:grad_thindisk}, we can see that adding this sample to that of panel 2 will act to pull the gradient upwards at all points. If we consider the leftmost column (panel 1) of Figure \ref{fig:grad_zmax}, we can see that restricting our data to having a small \zmax \ (the other constraint in Figure \ref{fig:grad_thindisk}) will act to pull the gradient for young stars downward.

In this way, Figures \ref{fig:grad_thindisk} and \ref{fig:grad_thickdisk} can be reconstructed qualitatively from Figures \ref{fig:grad_zmax} and \ref{fig:grad_chem}.

\subsubsection{The Choice of Galactic Parameters}

\begin{figure*}
	\includegraphics[width=\linewidth]{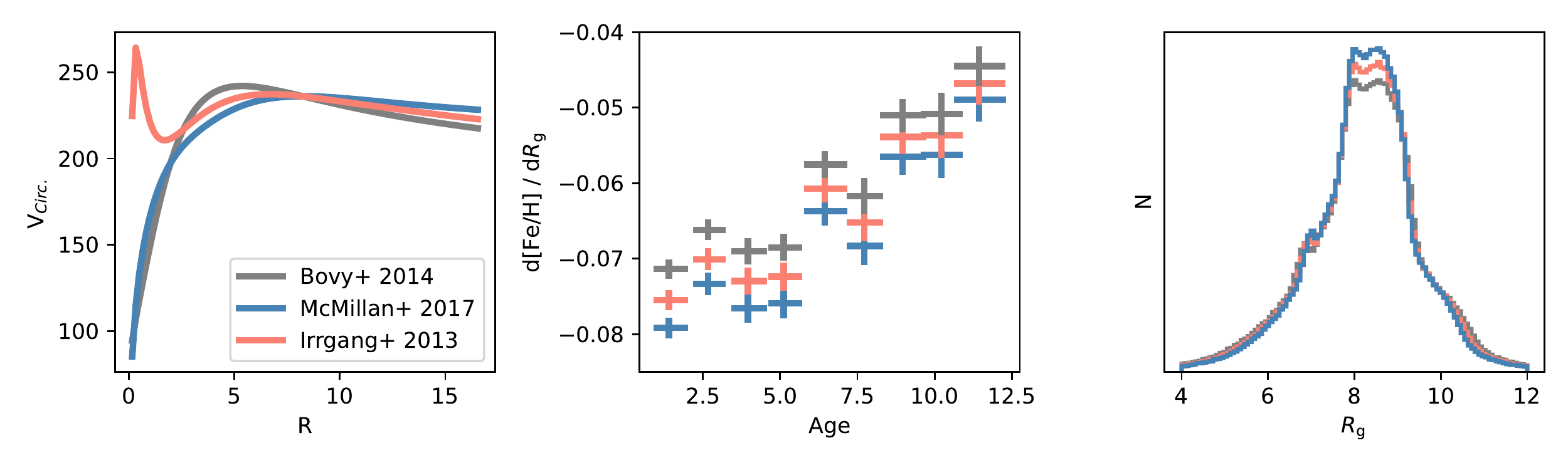}
	\caption{
          The effect of using various potentials when fitting our best thin disk sample (the sample shown in Figure \ref{fig:grad_thindisk}). We can see that the choice of potential does not dramatically change the results, but rather just shifts the metallicity gradient higher or lower in a systemic way.
        }
	\label{fig:potentials}
\end{figure*}

As described, our determination of \rguide \ is drawn directly from the rotation curve of our potential. As such, it is worth investigating the effects of different potentials on our results. We consider, in addition to the modified MWPotential2014 potential we have used for the main portion of this work, the potentials of \citet{irr2013} and \citet{mcm2017} in Figure \ref{fig:potentials}. The global trends in the fits of the metallicity gradient seem relatively unaffected by the choice of potential.

\begin{figure*}
	\includegraphics[width=\linewidth]{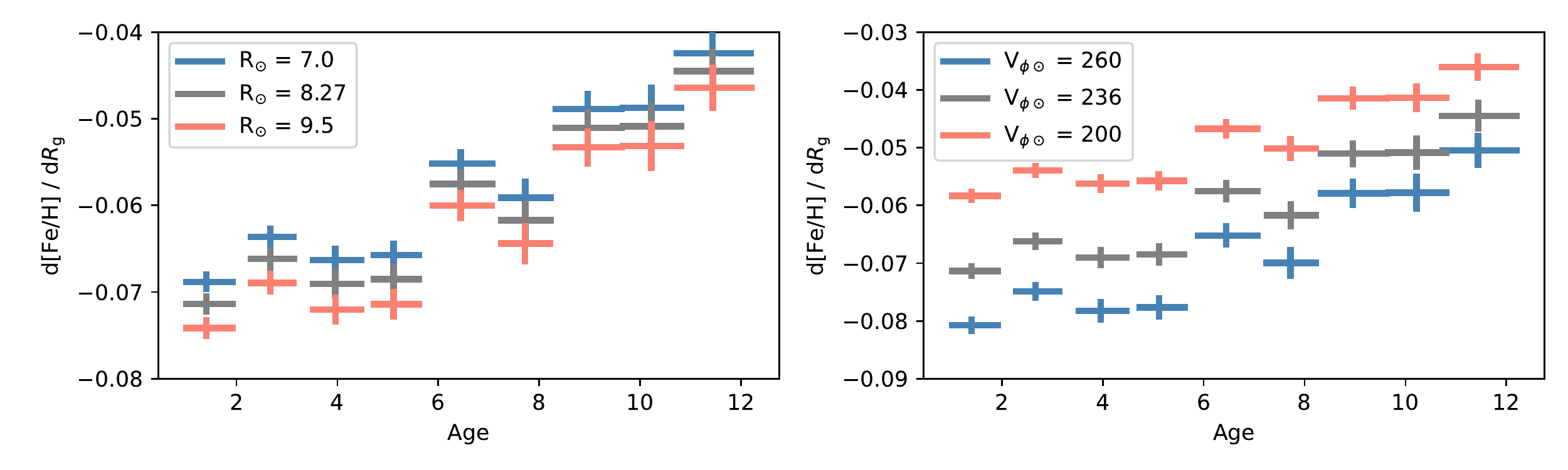}
	\caption{
          The effect of using various solar circle radius and rotation speed assumptions when fitting our best thin disk sample (the sample shown in Figure \ref{fig:grad_thindisk}). We can see that the choice can change our results outside of the allowance of the error-bars, and the slope of the fits seems strongly dependent on the rotational velocity assumption, but the global trend of the metallicity gradient decreasing as age increases remains for all assumptions considered here.
        }
	\label{fig:galpars}
\end{figure*}

While the choice of potential doesn't have much of an effect on the measured metallicity gradient, the choice of solar parameters -- that is the radius of the solar circle from the center of the Galaxy and local rotation speed -- can make a relatively larger difference. In Figure \ref{fig:galpars} we show the effect of changing these parameters. The gradient will shift upward and downward in magnitude for various choices of the solar radius, $R_{0}$. The general trend and slope of the gradient in age in our thin disk sample remains relatively unchanged. The choice of rotation speed can change the trend more dramatically, altering the slope of the metallicity gradient with age. In all of these examples, however, the pattern of younger stars having steeper gradients than older stars remains.

\subsubsection{More Reliable Age Values}\label{sec:reliable}

Age determinations are complicated. One particular weakness of the method which we use in this work, isochrone matching, is that stellar model isochrones tend to grow closer in the areas of the giant branch and the cooler portions of the main sequence. This causes isochrone age determinations to lose some of their discriminating power between ages in these areas, as the differences between models grows smaller compared to the typical stellar atmospheric parameter's errors.

Many works circumvent this issue by restricting their sample to the turn-off and subgiant branch areas of the color-magnitude diagram, since, in this area, isochrones differ more dramatically with age, making age determinations more reliable.

\begin{figure}
	\includegraphics[width=\linewidth]{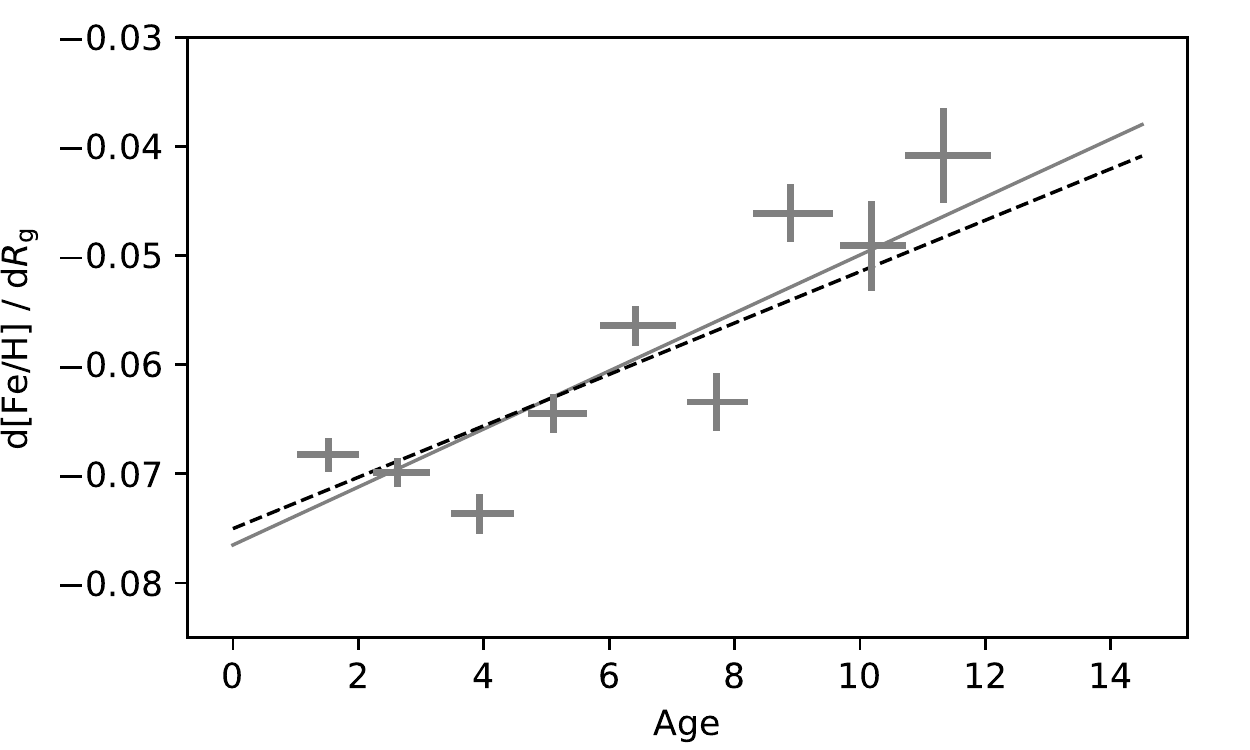}
	\caption{
          A zoom in on the top left panel of Figure \ref{fig:grad_thindisk} for our data set with age errors less than 10\%, and restricted to main-sequence turnoff and subgiant stars (grey crosses and fit line), showing little difference from the results using the dataset used in that Figure (shown here with a black dashed line) and the rest of the paper (which has age errors of 25\% and includes all stellar types).
        }
	\label{fig:high_prec_ages}
\end{figure}

We therefore construct a smaller, high quality sample from our data, consisting of 171,691 stars with 10\% or better age errors as well as: temperatures hotter than 5750 K, or log(g) $<$ 4 and hotter than 5250 K. This should remove the populations most likely to have confused age estimates. The results of fitting the metallicity gradients of this smaller, more reliable sample are shown in Figure \ref{fig:high_prec_ages}, and are similar to those found using our full sample.

\subsection{Explanations of the Evolving Metallicity Gradient}
\subsubsection{Thin Disk}
Star formation in the Milky Way is commonly thought of as growing in an inside-out (beginning in the central regions before expanding radially) and possibly upside-down (beginning with a vertically hot distribution which cools over time) fashion (for a more detailed explanation, see \citealt{bir2013}). In this scenario, one would expect the oldest stars to have steep radial metallicity gradients (as the central regions are enriched and little to no star formation occurs in the periphery) and the youngest stars to have shallower gradients (as star formation and enrichment spreads outward through the extent of the modern-day disk).

Instead we observe the opposite trend in our thin disk stars. This could be direct observational evidence of radial migration, the process by which stars' orbital radii may be shifted inward or outward in the galaxy and which results in a flattening of metallicity gradients over time.

Our analysis suggests that a single-parameter cut is not effective at selecting the population which gives the most constant flattening rate, since we see young age flattening and ``upticks'' which lead to ``U'' shapes in the d[Fe/H]/d\rguide \ vs age plots (the left and middle-left panels in Figures \ref{fig:grad_zmax} and \ref{fig:grad_chem}, as compared to Figure \ref{fig:grad_thindisk}). This behavior could be indicative of contamination from other populations. Using kinematic and chemistry cuts in tandem seems to isolate the gradient-flattening trend more effectively than either individual cut.

If these upticks are caused by a contamination from a distinct population such as the thick disk (which seems to have a higher, near-constant gradient of 0.02 dex kpc$^{-1}$), it poses an interesting conundrum in that we see evidence of this contamination most in the youngest age bins. It should not be an effect of noisy observations ``piling up'' in the edge (youngest) age-bins, since the trend extends past the edge of the isochrone grid to ages of 5 Gyr. That the chemical thick disk is not exclusively old is an idea that has drawn attention recently with growing numbers of observations of young, $\alpha$-enhanced stars; however, many of these observations imply mass transfer origins, so their atmospheres could appear ``young'' even though they are members of a kinematically older population.

The selection which makes the ``churning'' signal the strongest -- that is, the cut where the trend between d[Fe/H]/d\rguide \ is most monotonic with age -- is the cut based on orbital \zmax. This could be because of a ``provenance bias,'' or preference for migratory forces tend to favor (although not exclusively) stars on vertically cooler and less eccentric orbits, as mentioned in Section \ref{sec:zmax}.

\subsubsection{Thick Disk}
The chemical and orbital thick disk selections yield positive metallicity gradients which are similar for all ages. That the gradients are similar, regardless of age, implies that the population is well mixed in age. It is surprising that it is not well mixed in radius which would, presumably, result in a zero radial gradient. It is further surprising that the gradient is slightly positive. This might indicate that the chemical evolution of the thin and thick disks differ.

Positive gradients have been observed in external galaxies' thick disks \citep{cre2010}. In the Milky Way this has been observed indirectly in the velocities of thick disk stars as a function of metallicity (as in \citealt{lee2011}, \citealt{kor2013}, among others) where it is seen that more metal rich thick disk stars actually rotate more quickly (implying that we are observing them near pericenter) than lower metallicity thick disk stars.

Other studies, such as by \citet{nid2014}, find that the high-$\alpha$ stars in APOGEE have remarkably similar chemical distributions at all observational radii and heights from the plane, a result confirmed by \citet{hay2015}. \citet{loe2016} posit that this effect could arise in a simulation as a result of these stars being formed in a narrow radial range in a limited time-frame.

We suggest that these observations of no gradient in previous works studying high-$\alpha$ stars in \rpres \ could be consistent with our observation of a slightly positive gradient in \rguide \ for our ``thick disk'' selection if the data are blurred. It's worth mentioning that high-$\alpha$ stars are not necessarily synonymous with thick disk stars. We again note that selection functions make such comparisons more complex, especially when comparing the results of one spectroscopic survey to another.
  
It has been suggested that a positive d[Fe/H]/dR gradient may arise from the accretion of pristine gas into the center of the Galaxy at early epochs (\citealt{cre2010}, \citealt{cur2012}), and that it may be a natural occurrence during inside-out formation (for example: \citealt{sch2017ii}, \citealt{kaw2018}).

Our findings therefore are in agreement with other observations with respect to the positive gradients (for example \citealt{jia2018}, see also Table 1 of \citealt{tas2016}). That this gradient is similar at all ages is a result which should help constrain the formation history of the Milky Way's thick disk.

\section{Conclusions}\label{sec:conclusions}

We have calculated the ages, phase-space coordinates, and orbits for $\sim$4 million stars in LAMOST using the atmospheric parameters of the LAMOST pipeline, $\alpha$ abundances from \citet{xia2019}, astrometry from Gaia eDR3 \citep{gai2020}, and distances from \citep{bai2020}.

Using a subset of about 1.3 million stars with relatively accurate ages, we have analyzed the metallicity gradient as a function of Galactic radius and stellar age in various slices designed to select the chemical and physical thin and thick disks. We find that:

\begin{itemize}
\item{The stars of the orbital thin disk, defined as \zmax \ $<$ 250 pc, display a monotonically flattening metallicity gradient, with the youngest stars having the steepest metallicity gradients and the older stars having shallower gradients.}
\item{The chemical thin disk, defined by [$\alpha$/Fe] $<$ 0.15, [$\alpha$/Fe] $<$ -0.2$\cdot$[Fe/H] $+$ 0.1, behaves similarly to the orbital thin disk, with a shallower metallicity gradient for old stars, and a steeper gradient for young stars.}
\item{The thin disk defined by both of these cuts has a metallicity gradient of -0.075 dex kpc$^{-1}$ for the youngest stars, and this gradient gradually grows shallower with age at a rate of 0.003 dex kpc$^{-1}$ Gyr$^{-1}$.}
\item{The orbital thick disk, defined as \zmax \ $>$ 2 kpc, and the chemical thick disk, defined by [$\alpha$/Fe] $>$ 0.15 or [$\alpha$/Fe] $>$ -0.2$\cdot$[Fe/H] $+$ 0.1, seem to have positive metallicity gradients near 0.02 dex kpc$^{-1}$ for all ages. The reason for this is still not well understood.}
\end{itemize}

\section{Acknowledgements}

We thank the referee for their comments which helped to improve the quality of this work.

We thank Xiaoting Fu for helpful discussions related to stellar ages, isochrones, and cluster comparisons; Maosheng Xiang for helpful discussion related to ages and alpha abundances, and for comments on an early draft; Ji Li for helpful discussion about $\alpha$ abundance determinations; Jerry Sellwood and Victor Debattista for helpful comments on an early draft.

We thank the developers and maintainers of the following software libraries which were used in this work: Topcat \citep{tay2005},  NumPy \citep{van2011}, SciPy \citep{jon2001}, AstroPy \citep{ast2013}, matplotlib  \citep{hun2007}, scikit-learn \citep{ped2011}, IPython \citep{per2007}, and Python.

The research presented here is partially supported by the National Key R\&D Program of China under grant No. 2018YFA0404501; by the National Natural Science Foundation of China under grant Nos. 12025302, 11773052, 11761131016; by the ``111'' Project of the Ministry of Education under grant No. B20019. ZYL is supported by a Shanghai Natural Science Research Grant (21ZR1430600). This work made use of the Gravity Supercomputer at the Department of Astronomy, Shanghai Jiao Tong University, and the facilities of the Center for High Performance Computing at Shanghai Astronomical Observatory.

JJV gratefully acknowledges the support of the Chinese Academy of Sciences President's International Fellowship Initiative.

Guoshoujing Telescope (the Large Sky Area Multi-Object Fiber Spectroscopic Telescope LAMOST) is a National Major Scientific Project built by the Chinese Academy of Sciences. Funding for the project has been provided by the National Development and Reform Commission. LAMOST is operated and managed by the National Astronomical Observatories, Chinese Academy of Sciences.

This work has made use of data from the European Space Agency (ESA) mission {\it Gaia} (\url{https://www.cosmos.esa.int/gaia}), processed by the {\it Gaia} Data Processing and Analysis Consortium (DPAC, \url{https://www.cosmos.esa.int/web/gaia/dpac/consortium}). Funding for the DPAC has been provided by national institutions, in particular the institutions participating in the {\it Gaia} Multilateral Agreement.


\begin{thebibliography}{56}

\bibitem[Astropy Collaboration et al.(2013)]{ast2013} Astropy Collaboration, Robitaille, T.~P., Tollerud, E.~J., et al.\ 2013, \aap, 558, A33

\bibitem[Anders et al.(2017)]{and2017} Anders, F., Chiappini, C., Minchev, I., et al.\ 2017, \aap, 600, A70. doi:10.1051/0004-6361/201629363

\bibitem[Aumer et al.(2016)]{aum2016} Aumer, M., Binney, J., \& Sch{\"o}nrich, R.\ 2016, \mnras, 462, 1697. doi:10.1093/mnras/stw1639

\bibitem[Bailer-Jones et al.(2020)]{bai2020} Bailer-Jones, C.~A.~L., Rybizki, J., Fouesneau, M., et al.\ 2020, arXiv:2012.05220

\bibitem[Barnes et al.(2016)]{bar2016} Barnes, S.~A., Weingrill, J., Fritzewski, D., et al.\ 2016, \apj, 823, 16. doi:10.3847/0004-637X/823/1/16

\bibitem[Bird et al.(2013)]{bir2013} Bird, J.~C., Kazantzidis, S., Weinberg, D.~H., et al.\ 2013, \apj, 773, 43. doi:10.1088/0004-637X/773/1/43

\bibitem[Boggs \& Rogers(1990)]{bog1990} P. T. Boggs and J. E. Rogers, “Orthogonal Distance Regression,” in “Statistical analysis of measurement error models and applications: proceedings of the AMS-IMS-SIAM joint summer research conference held June 10-16, 1989,” Contemporary Mathematics, vol. 112, pg. 186, 1990.

\bibitem[Borucki et al.(2010)]{bor2010} Borucki, W.~J., Koch, D., Basri, G., et al.\ 2010, Science, 327, 977. doi:10.1126/science.1185402

\bibitem[Bovy(2015)]{bov2015} Bovy, J.\ 2015, \apjs, 216, 29

\bibitem[Bovy et al.(2016)]{bov2016} Bovy, J., Rix, H.-W., Green, G.~M., et al.\ 2016, \apj, 818, 130. doi:10.3847/0004-637X/818/2/130

\bibitem[Bressan et al.(2012)]{bre2012} Bressan, A., Marigo, P., Girardi, L., et al.\ 2012, \mnras, 427, 127. doi:10.1111/j.1365-2966.2012.21948.x

\bibitem[Cantat-Gaudin et al.(2018)]{can2018} Cantat-Gaudin, T., Jordi, C., Vallenari, A., et al.\ 2018, \aap, 618, A93. doi:10.1051/0004-6361/201833476

\bibitem[Casagrande et al.(2011)]{cas2011} Casagrande, L., Sch{\"o}nrich, R., Asplund, M., et al.\ 2011, \aap, 530, A138. doi:10.1051/0004-6361/201016276

\bibitem[Chen \& Zhao(2020)]{che2020} Chen, Y.~Q. \& Zhao, G.\ 2020, \mnras, 495, 2673. doi:10.1093/mnras/staa1079

\bibitem[Chiappini et al.(2015)]{chi2015} Chiappini, C., Anders, F., Rodrigues, T.~S., et al.\ 2015, \aap, 576, L12. doi:10.1051/0004-6361/201525865

\bibitem[Cresci et al.(2010)]{cre2010} Cresci, G., Mannucci, F., Maiolino, R., et al.\ 2010, \nat, 467, 811. doi:10.1038/nature09451

\bibitem[Cui et al.(2012)]{cui2012} Cui, X.-Q., Zhao, Y.-H., Chu, Y.-Q., et al.\ 2012, Research in Astronomy and Astrophysics, 12, 1197. doi:10.1088/1674-4527/12/9/003
  
\bibitem[Curir et al.(2012)]{cur2012} Curir, A., Lattanzi, M.~G., Spagna, A., et al.\ 2012, \aap, 545, A133. doi:10.1051/0004-6361/201118558

\bibitem[De Silva et al.(2015)]{des2015} De Silva, G.~M., Freeman, K.~C., Bland-Hawthorn, J., et al.\ 2015, \mnras, 449, 2604. doi:10.1093/mnras/stv327

\bibitem[Dehnen \& Binney(1998)]{deh1998} Dehnen, W. \& Binney, J.~J.\ 1998, \mnras, 298, 387. doi:10.1046/j.1365-8711.1998.01600.x

\bibitem[Demarque et al.(2004)]{dem2004} Demarque, P., Woo, J.-H., Kim, Y.-C., et al.\ 2004, \apjs, 155, 667. doi:10.1086/424966

\bibitem[Deng et al.(2012)]{den2012} Deng, L.-C., Newberg, H.~J., Liu, C., et al.\ 2012, Research in Astronomy and Astrophysics, 12, 735. doi:10.1088/1674-4527/12/7/003

\bibitem[Dias et al.(2021)]{dia2021} Dias, W.~S., Monteiro, H., Moitinho, A., et al.\ 2021, \mnras, 504, 356. doi:10.1093/mnras/stab770

\bibitem[Dotter et al.(2007)]{dot2007} Dotter, A., Chaboyer, B., Jevremovi{\'c}, D., et al.\ 2007, \aj, 134, 376 

\bibitem[Dotter et al.(2008)]{dot2008} Dotter, A., Chaboyer, B., Jevremovi{\'c}, D., et al.\ 2008, \apjs, 178, 89 

\bibitem[Drimmel et al.(2003)]{dri2003} Drimmel, R., Cabrera-Lavers, A., \& L{\'o}pez-Corredoira, M.\ 2003, \aap, 409, 205. doi:10.1051/0004-6361:20031070

\bibitem[Edvardsson et al.(1993)]{edv1993} Edvardsson, B., Andersen, J., Gustafsson, B., et al.\ 1993, \aap, 500, 391

\bibitem[Frankel et al.(2020)]{fra2020} Frankel, N., Sanders, J., Ting, Y.-S., et al.\ 2020, \apj, 896, 15. doi:10.3847/1538-4357/ab910c

\bibitem[Fu et al.(2020)]{fu2020} Fu, J.-N., Cat, P.~D., Zong, W., et al.\ 2020, Research in Astronomy and Astrophysics, 20, 167. doi:10.1088/1674-4527/20/10/167

%DR3
\bibitem[Gaia Collaboration et al.(2020)]{gai2020} Gaia Collaboration, Brown, A.~G.~A., Vallenari, A., et al.\ 2020, arXiv:2012.01533

\bibitem[Gibson et al.(2013)]{gib2013} Gibson, B.~K., Pilkington, K., Brook, C.~B., et al.\ 2013, \aap, 554, A47. doi:10.1051/0004-6361/201321239

\bibitem[Gilmore \& Reid(1983)]{gil1983} Gilmore, G. \& Reid, N.\ 1983, \mnras, 202, 1025. doi:10.1093/mnras/202.4.1025

\bibitem[Gratton et al.(1996)]{gra1996} Gratton, R., Carretta, E., Matteucci, F., et al.\ 1996, Formation of the Galactic Halo...Inside and Out, 92, 307
  
\bibitem[Green et al.(2015)]{gre2015} Green, G.~M., Schlafly, E.~F., Finkbeiner, D.~P., et al.\ 2015, \apj, 810, 25. doi:10.1088/0004-637X/810/1/25

\bibitem[Hayden et al.(2015)]{hay2015} Hayden, M.~R., Bovy, J., Holtzman, J.~A., et al.\ 2015, \apj, 808, 132. doi:10.1088/0004-637X/808/2/132

\bibitem[Hekker \& Johnson(2019)]{hek2019} Hekker, S. \& Johnson, J.~A.\ 2019, \mnras, 487, 4343. doi:10.1093/mnras/stz1554

\bibitem[Holmberg et al.(2009)]{hol2009} Holmberg, J., Nordstr{\"o}m, B., \& Andersen, J.\ 2009, \aap, 501, 941. doi:10.1051/0004-6361/200811191

\bibitem[Huang et al.(2015)]{hua2015} Huang, Y., Liu, X.-W., Yuan, H.-B., et al.\ 2015, \mnras, 454, 2863. doi:10.1093/mnras/stv1991

\bibitem[Hunter(2007)]{hun2007} Hunter J. D., 2007, Computing In Science \& Engineering, 9, 90

\bibitem[Irrgang et al.(2013)]{irr2013} Irrgang, A., Wilcox, B., Tucker, E., et al.\ 2013, \aap, 549, A137. doi:10.1051/0004-6361/201220540

\bibitem[Jia et al.(2018)]{jia2018} Jia, Y., Chen, Y., Zhao, G., et al.\ 2018, \apj, 863, 93. doi:10.3847/1538-4357/aad3bb

\bibitem[Jofr{\'e} et al.(2016)]{jof2016} Jofr{\'e}, P., Jorissen, A., Van Eck, S., et al.\ 2016, \aap, 595, A60. doi:10.1051/0004-6361/201629356

\bibitem[Johnson \& Soderblom(1987)]{joh1987} Johnson, D.~R.~H., \& Soderblom, D.~R.\ 1987, \aj, 93, 864

\bibitem[Jones et al.(2001)]{jon2001}Jones E., Oliphant T., Peterson P., et al., 2001, SciPy: Open source scientific tools for Python, http://www.scipy.org/

\bibitem[J{\o}rgensen \& Lindegren(2005)]{jor2005} J{\o}rgensen, B.~R. \& Lindegren, L.\ 2005, \aap, 436, 127. doi:10.1051/0004-6361:20042185

\bibitem[Juri{\'c} et al.(2008)]{jur2008} Juri{\'c}, M., Ivezi{\'c}, {\v{Z}}., Brooks, A., et al.\ 2008, \apj, 673, 864. doi:10.1086/523619
  
\bibitem[Kawata et al.(2018)]{kaw2018} Kawata, D., Allende Prieto, C., Brook, C.~B., et al.\ 2018, \mnras, 473, 867. doi:10.1093/mnras/stx2464

\bibitem[Kim et al.(2002)]{kim2002} Kim, Y.-C., Demarque, P., Yi, S.~K., et al.\ 2002, \apjs, 143, 499. doi:10.1086/343041

\bibitem[Kordopatis et al.(2013)]{kor2013} Kordopatis, G., Gilmore, G., Wyse, R.~F.~G., et al.\ 2013, \mnras, 436, 3231. doi:10.1093/mnras/stt1804

\bibitem[Kroupa(2001)]{kro2001} Kroupa, P.\ 2001, \mnras, 322, 231

\bibitem[Lacey(1984)]{lac1984} Lacey, C.~G.\ 1984, \mnras, 208, 687. doi:10.1093/mnras/208.4.687
  
\bibitem[Lee et al.(2011)]{lee2011} Lee, Y.~S., Beers, T.~C., An, D., et al.\ 2011, \apj, 738, 187. doi:10.1088/0004-637X/738/2/187

\bibitem[Liu et al.(2014)]{liu2014} Liu, X.-W., Yuan, H.-B., Huo, Z.-Y., et al.\ 2014, Setting the scene for Gaia and LAMOST, 298, 310. doi:10.1017/S1743921313006510
  
\bibitem[Loebman et al.(2016)]{loe2016} Loebman, S.~R., Debattista, V.~P., Nidever, D.~L., et al.\ 2016, \apjl, 818, L6. doi:10.3847/2041-8205/818/1/L6

\bibitem[Luo et al.(2015)]{luo2015} Luo, A.-L., Zhao, Y.-H., Zhao, G., et al.\ 2015, Research in Astronomy and Astrophysics, 15, 1095

\bibitem[Mackereth et al.(2017)]{mac2017} Mackereth, J.~T., Bovy, J., Schiavon, R.~P., et al.\ 2017, \mnras, 471, 3057. doi:10.1093/mnras/stx1774

\bibitem[Mackereth et al.(2019)]{mac2019} Mackereth, J.~T., Bovy, J., Leung, H.~W., et al.\ 2019, \mnras, 489, 176. doi:10.1093/mnras/stz1521

\bibitem[Majewski et al.(2017)]{maj2017} Majewski, S.~R., Schiavon, R.~P., Frinchaboy, P.~M., et al.\ 2017, \aj, 154, 94. doi:10.3847/1538-3881/aa784d

\bibitem[Marshall et al.(2006)]{mar2006} Marshall, D.~J., Robin, A.~C., Reyl{\'e}, C., et al.\ 2006, \aap, 453, 635. doi:10.1051/0004-6361:20053842

\bibitem[Martig et al.(2015)]{mar2015} Martig, M., Rix, H.-W., Silva Aguirre, V., et al.\ 2015, \mnras, 451, 2230. doi:10.1093/mnras/stv1071

\bibitem[Matsuno et al.(2018)]{mat2018} Matsuno, T., Yong, D., Aoki, W., et al.\ 2018, \apj, 860, 49. doi:10.3847/1538-4357/aac019

\bibitem[Matteucci \& Francois(1989)]{mat1989} Matteucci, F. \& Francois, P.\ 1989, \mnras, 239, 885. doi:10.1093/mnras/239.3.885
  
\bibitem[McMillan(2011)]{mcm2011} McMillan, P.~J.\ 2011, \mnras, 414, 2446

\bibitem[McMillan(2017)]{mcm2017} McMillan, P.~J.\ 2017, \mnras, 465, 76. doi:10.1093/mnras/stw2759

\bibitem[Mikkola et al.(2020)]{mik2020} Mikkola, D., McMillan, P.~J., \& Hobbs, D.\ 2020, \mnras, 495, 3295. doi:10.1093/mnras/staa1223

\bibitem[Minchev \& Famaey(2010)]{min2010} Minchev, I. \& Famaey, B.\ 2010, \apj, 722, 112. doi:10.1088/0004-637X/722/1/112

\bibitem[Minchev et al.(2018)]{min2018} Minchev, I., Anders, F., Recio-Blanco, A., et al.\ 2018, \mnras, 481, 1645. doi:10.1093/mnras/sty2033

\bibitem[Miyamoto \& Nagai(1975)]{miy1975} Miyamoto, M., \& Nagai, R.\ 1975, \pasj, 27, 533

\bibitem[Navarro et al.(1997)]{nav1997} Navarro, J.~F., Frenk, C.~S., \& White, S.~D.~M.\ 1997, \apj, 490, 493

\bibitem[Ness et al.(2015)]{nes2015} Ness, M., Hogg, D.~W., Rix, H.-W., et al.\ 2015, \apj, 808, 16. doi:10.1088/0004-637X/808/1/16

\bibitem[Nidever et al.(2014)]{nid2014} Nidever, D.~L., Bovy, J., Bird, J.~C., et al.\ 2014, \apj, 796, 38. doi:10.1088/0004-637X/796/1/38

\bibitem[Nissen(2004)]{nis2004} Nissen, P.~E.\ 2004, Origin and Evolution of the Elements, 154

\bibitem[Nordstr{\"o}m et al.(2004)]{nor2004} Nordstr{\"o}m, B., Mayor, M., Andersen, J., et al.\ 2004, \aap, 418, 989. doi:10.1051/0004-6361:20035959
  
\bibitem[Ojha et al.(1996)]{ojh1996} Ojha, D.~K., Bienayme, O., Robin, A.~C., et al.\ 1996, \aap, 311, 456

\bibitem[Oort(1922)]{oor1922} Oort, J.~H.\ 1922, \bain, 1, 133

\bibitem[{\"O}nal Ta{\c{s}} et al.(2016)]{tas2016} {\"O}nal Ta{\c{s}}, {\"O}., Bilir, S., Seabroke, G.~M., et al.\ 2016, \pasa, 33, e044. doi:10.1017/pasa.2016.33

\bibitem[Pedregosa(2011)]{ped2011} Pedregosa F., 2011, Journal of Machine Learning Research, 12, 2825

\bibitem[Perez \& Granger(2007)]{per2007} Perez, F. \& Granger, B.~E.\ 2007, Computing in Science and Engineering, 9, 21. doi:10.1109/MCSE.2007.53

\bibitem[Pilkington et al.(2012)]{pil2012} Pilkington, K., Few, C.~G., Gibson, B.~K., et al.\ 2012, \aap, 540, A56. doi:10.1051/0004-6361/201117466

\bibitem[Pinsonneault et al.(2018)]{pin2018} Pinsonneault, M.~H., Elsworth, Y.~P., Tayar, J., et al.\ 2018, \apjs, 239, 32. doi:10.3847/1538-4365/aaebfd

\bibitem[Ro{\v{s}}kar et al.(2012)]{ros2012} Ro{\v{s}}kar, R., Debattista, V.~P., Quinn, T.~R., et al.\ 2012, \mnras, 426, 2089. doi:10.1111/j.1365-2966.2012.21860.x

\bibitem[Ro{\v{s}}kar et al.(2008)]{ros2008} Ro{\v{s}}kar, R., Debattista, V.~P., Quinn, T.~R., et al.\ 2008, \apjl, 684, L79. doi:10.1086/592231

\bibitem[Sanders \& Das(2018)]{san2018} Sanders, J.~L. \& Das, P.\ 2018, \mnras, 481, 4093. doi:10.1093/mnras/sty2490

\bibitem[Sch{\"o}nrich \& Binney(2009)]{sch2009} Sch{\"o}nrich, R. \& Binney, J.\ 2009, \mnras, 396, 203. doi:10.1111/j.1365-2966.2009.14750.x

\bibitem[Sch{\"o}nrich(2012)]{sch2012} Sch{\"o}nrich, R.\ 2012, \mnras, 427, 274

\bibitem[Sch{\"o}nrich \& Aumer(2017)]{sch2017} Sch{\"o}nrich, R., \& Aumer, M.\ 2017, arXiv:1704.01333

\bibitem[Sch{\"o}nrich \& McMillan(2017)]{sch2017ii} Sch{\"o}nrich, R. \& McMillan, P.~J.\ 2017, \mnras, 467, 1154. doi:10.1093/mnras/stx093

\bibitem[Sellwood \& Binney(2002)]{sel2002} Sellwood, J.~A. \& Binney, J.~J.\ 2002, \mnras, 336, 785. doi:10.1046/j.1365-8711.2002.05806.x

\bibitem[Sellwood(2014)]{sel2014} Sellwood, J.~A.\ 2014, Reviews of Modern Physics, 86, 1. doi:10.1103/RevModPhys.86.1

\bibitem[Soderblom(2010)]{sod2010} Soderblom, D.~R.\ 2010, \araa, 48, 581. doi:10.1146/annurev-astro-081309-130806

\bibitem[Solway et al.(2012)]{sol2012} Solway, M., Sellwood, J.~A., \& Sch{\"o}nrich, R.\ 2012, \mnras, 422, 1363. doi:10.1111/j.1365-2966.2012.20712.x

\bibitem[Spitzer \& Schwarzschild(1951)]{spi1951} Spitzer, L. \& Schwarzschild, M.\ 1951, \apj, 114, 385. doi:10.1086/145478
  
\bibitem[Str{\"o}mberg(1924)]{str1924} Str{\"o}mberg, G.\ 1924, \apj, 59, 228. doi:10.1086/142813

\bibitem[Str{\"o}mberg(1946)]{str1946} Str{\"o}mberg, G.\ 1946, \apj, 104, 12. doi:10.1086/144830
  
\bibitem[Sun et al.(2020)]{sun2020} Sun, W.-X., Huang, Y., Wang, H.-F., et al.\ 2020, \apj, 903, 12. doi:10.3847/1538-4357/abb1b7
  
\bibitem[Taylor(2005)]{tay2005} Taylor, M.~B.\ 2005, Astronomical Data Analysis Software and Systems XIV, 347, 29

\bibitem[Ting et al.(2019)]{tin2019} Ting, Y.-S., Conroy, C., Rix, H.-W., et al.\ 2019, \apj, 879, 69. doi:10.3847/1538-4357/ab2331

\bibitem[van der Walt et al.(2011)]{van2011} van der Walt S., Colbert S. C., Varoquaux G., 2011, Computing in Science \& Engineering, 13

\bibitem[Vera-Ciro et al.(2014)]{ver2014} Vera-Ciro, C., D'Onghia, E., Navarro, J., et al.\ 2014, \apj, 794, 173. doi:10.1088/0004-637X/794/2/173

\bibitem[Vera-Ciro et al.(2016)]{ver2016} Vera-Ciro, C., D'Onghia, E., \& Navarro, J.~F.\ 2016, \apj, 833, 42. doi:10.3847/1538-4357/833/1/42

\bibitem[Vickers \& Smith(2018)]{vic2018} Vickers, J.~J. \& Smith, M.~C.\ 2018, \apj, 860, 91. doi:10.3847/1538-4357/aac323

\bibitem[Vincenzo et al.(2021)]{vin2021} Vincenzo, F., Weinberg, D.~H., Miglio, A., et al.\ 2021, arXiv:2101.04488

\bibitem[Wang et al.(2019)]{wan2019} Wang, C., Liu, X.-W., Xiang, M.-S., et al.\ 2019, \mnras, 482, 2189. doi:10.1093/mnras/sty2797

\bibitem[Wu et al.(2014)]{wu2014} Wu, Y., Du, B., Luo, A., et al.\ 2014, Statistical Challenges in 21st Century Cosmology, 306, 340. doi:10.1017/S1743921314010825

\bibitem[Wyse \& Gilmore(1986)]{wys1986} Wyse, R.~F.~G. \& Gilmore, G.\ 1986, \aj, 91, 855. doi:10.1086/114064

\bibitem[Xiang et al.(2015)]{xia2015} Xiang, M.-S., Liu, X.-W., Yuan, H.-B., et al.\ 2015, Research in Astronomy and Astrophysics, 15, 1209. doi:10.1088/1674-4527/15/8/009

\bibitem[Xiang et al.(2017)]{xia2017} Xiang, M., Liu, X., Shi, J., et al.\ 2017, \apjs, 232, 2. doi:10.3847/1538-4365/aa80e4

\bibitem[Xiang et al.(2019)]{xia2019} Xiang, M., Ting, Y.-S., Rix, H.-W., et al.\ 2019, \apjs, 245, 34. doi:10.3847/1538-4365/ab5364

\bibitem[Yi et al.(2001)]{yi2001} Yi, S., Demarque, P., Kim, Y.-C., et al.\ 2001, \apjs, 136, 417. doi:10.1086/321795

\bibitem[Yi et al.(2003)]{yi2003} Yi, S.~K., Kim, Y.-C., \& Demarque, P.\ 2003, \apjs, 144, 259. doi:10.1086/345101

\bibitem[Yong et al.(2016)]{yon2016} Yong, D., Casagrande, L., Venn, K.~A., et al.\ 2016, \mnras, 459, 487. doi:10.1093/mnras/stw676

\bibitem[Yoshii(1982)]{yos1982} Yoshii, Y.\ 1982, \pasj, 34, 365

\bibitem[Yu et al.(2012)]{yu2012} Yu, J., Sellwood, J.~A., Pryor, C., et al.\ 2012, \apj, 754, 124. doi:10.1088/0004-637X/754/2/124

\bibitem[Yuan et al.(2015)]{yua2015} Yuan, H.-B., Liu, X.-W., Huo, Z.-Y., et al.\ 2015, \mnras, 448, 855. doi:10.1093/mnras/stu2723

\bibitem[Zhang et al.(2021)]{zha2021} Zhang, H., Chen, Y., \& Zhao, G.\ 2021, arXiv:2106.12841    

\end{thebibliography}
\end{document}